\documentclass[12pt]{article}
\usepackage{amsmath}
\usepackage{graphicx}
\usepackage{enumerate}
\usepackage{natbib}
\usepackage{url} 

\usepackage{amsmath, amssymb,amsthm, amscd, amsfonts}
\usepackage{wrapfig, graphicx, caption, subcaption}
\usepackage[space]{grffile}

\usepackage[table]{xcolor}

\usepackage{tikz}
\usetikzlibrary{arrows}

\newcommand{\blind}{1}

\graphicspath{{./figures/}}

\usepackage{fancyhdr}
\usepackage[letterpaper, left=2.5cm, right=2.5cm, top=2.5cm,
bottom=2.5cm,dvips]{geometry}
\usepackage{url,verbatim}
\usepackage[normalem]{ulem}
\usepackage[colorlinks,citecolor=blue,urlcolor=blue]{hyperref}
\usepackage{color}
\usepackage[ruled,  lined,  commentsnumbered]{algorithm2e}
\usepackage{marginnote} 
\renewcommand{\marginpar}[2][]{}

\theoremstyle{plain}
\newtheorem{theorem}{Theorem}[section]

\newtheorem{proposition}[theorem]{Proposition}

\theoremstyle{definition}
\newtheorem{defn}[theorem]{Definition}

\newtheorem{remark}[theorem]{Remark}

\definecolor{forestgreen}{rgb}{0,.72,0} 
\definecolor{brickred}{rgb}{.72,0,0}
\definecolor{darkcerulean}{rgb}{0.03, 0.27, 0.49}
\newcommand{\sonja}[1]{\textcolor{cyan}{{\it S: #1}}}

\newcommand{\R}{\mathbb{R}}

\newcommand{\F}{\mathcal F}
\newcommand{\cH}{H}
\newcommand{\M}{\mathcal M}

\newcommand{\Z}{\mathbb Z}
\newcommand{\cB}{\mathcal B}
\newcommand{\cR}{\mathcal R}
\newcommand{\G}{\mathcal G}

\DeclareMathOperator{\Unif}{Unif}

\renewcommand{\vec}[1]{\mathbf{#1}}

\begin{document}

\bibliographystyle{natbib}


\if1\blind
{
  \title{\bf Goodness of fit for log-linear ERGMs}
  \author{Elizabeth Gross\thanks{
    EG gratefully acknowledges support by National Science Foundation  under grant DMS-1620109, and SP support by DOE award \#1010629 and the Simons Foundation Gift \#854770.}\hspace{.2cm}\\
     University of Hawai'i at Mānoa\\
    and \\
    Sonja Petrovi\'{c}\\
    Illinois Institute of Technology\\
    and \\
    Despina Stasi \\
    Illinois Institute of Technology}
  \maketitle
} \fi

\if0\blind
{
  \bigskip
  \bigskip
  \bigskip
  \begin{center}
    {\LARGE\bf {Goodness of fit for log-linear ERGMs}
\end{center}
  \medskip
} \fi

\begin{abstract} 
Many popular models from the networks literature  can be viewed through  a common lens of contingency tables on network dyads, resulting in 
 \emph{log-linear ERGMs}:  
 exponential family models for random graphs whose sufficient statistics  are linear on the dyads. 
We propose a new model in this family, the \emph{$p_1$-SBM}, which combines node and group effects common in network formation mechanisms. In particular, it is a generalization of several well-known ERGMs including the stochastic blockmodel for undirected graphs with known block assignment, the degree-corrected version of it, and the directed $p_1$ model without group structure. 

We frame the problem of testing model fit for the log-linear ERGM class through 
 an exact conditional test whose $p$-value can be approximated efficiently in networks of  both small and moderately large sizes.  The sampling methods we build rely on a dynamic adaptation of Markov bases. We use quick estimation algorithms adapted from the contingency table literature and effective sampling methods rooted in graph theory and algebraic statistics. 
The performance and scalability of the method  is demonstrated on two data sets from biology: the connectome of \emph{C. elegans} and the interactome of  \emph{Arabidopsis thaliana}.  These two networks---a 
 network and a protein-protein interaction network---have been popular examples in the network science literature. Our work provides a model-based approach to studying them. 
\end{abstract}


\noindent%
{\it Keywords:}  exponential random graph models, networks, log-linear models, model goodness-of-fit, sparse contingency tables, Markov basis


\section{Introduction}
  Data in the form of graphs are common in many biological contexts.  In systems biology, graphs are used to record protein-protein interactions. For example, in \cite{AIMC2011}, the authors construct a network from 5664 experimentally observed interactions between 2661 proteins from \emph{Arabidopsis thaliana}, a small flowering plant used as a model organism.  In neuroscience, graphs are used to record synaptic contacts between neurons, as in the gap junction and chemical synapse networks constructed for \emph{C. elegans} in \cite{WattsStrogatz98}; see also a recent study of these networks in \cite{Varshneyetal11}. 
  Both protein-protein interaction networks and neuronal networks have been used as examples of scale-free networks, that is, networks whose degree distribution follows a power law (\cite{Jeong2000}, \cite{Tan2005}, \cite{Wagner2001}, \cite{Alb2005}, \cite{Clauset09}, \cite{Multilayer16}). Descriptive statistics have been used to study these networks, suggesting a  degree-based edge formation mechanisms in these data. However, both of these networks have yet to be rigorously analyzed using a model-fitting approach. 
 Part of the difficulty in studying such networks within a model-based setting is that statistical theory regarding fitting random graph models is still in development, since it poses several challenging combinatorial and algorithmic problems.

The present manuscript adds a chapter to the 40-year-long story on random graph models and the   difficult question of testing model fit.  
Even for exponential family random graph models (ERGMs), 
general quantitative methods for goodness-of-fit testing are still a challenge; see for example \citet[\S 2.3.4]{KolaczykBook2017}. 
Looking back at the literature of the past four decades reveals that implementing goodness-of-fit methods based on conditional exact tests is feasible,  especially if we begin by restricting to ERGMs that can be interpreted as log-linear models on contingency tables.

  Such models belong to the class of dyad-independent models in the network modeling literature. While this class of models is a restricted class that does not allow for modeling transitivity, they have found many applications from political blog networks \cite{Lei16}  to financial networks \cite{barucca2016disentangling} to biological networks \cite{cabreros2016detecting}, \cite{pavlovic2014stochastic}. The class includes, as we discuss in Section~\ref{sec:knownmodels}, undirected and directed degree-based models, the stochastic blockmodel (SBM), the degree-corrected version of the  SBM, and all of the above which include node covariate information.  The stochastic block model and its variants are perhaps the most broadly used due to their connection to clustering (for reviews see \cite{funke2019stochastic} and \cite{lee2019review}).  While in this paper we assume known block assignments, in most applications to real data, the block assignments are not observed and fitting includes the challenge of inferring both block assignments and parameter values simultaneously. However, due to their applications to clustering, as argued in \cite{lee2019review}, goodness-of-fit tests for stochastic block models with known block assignments are needed for selection and diagnostics during the clustering process.  Furthermore, \cite{karwa2016exact} provides a methodology to extend the exact test for any stochastic block model to a latent-block stochastic block model.

The following brief historical overview will further illuminate the motivation for our work. 
The connection between random graph models and log-linear models was made concretely in \cite{FW81}, a comment on \cite{HL81} that introduced the $p_1$ model, a degree-based model for directed networks. The  comment paper re-interpreted the model in log-linear form by organizing the network data into a four-dimensional table, thus drawing  a natural connection to contingency tables, with each cell  representing a classification of a dyad according to the particular state in the observed network. 
\citet{FW81}  proved the equivalence of the network and table models, linking random graphs to the exciting developments in categorical data analysis that were taking place at that time and, consequently, to the fitting methods developed over the years  and discussed below. 
The advantage of log-linear model representation to parameter estimation via iterative proportional scaling algorithms  was made clear,    even if it required a slight redundancy in the data representation. 
The connection inspired further work:  during the same decade, \citet{FienbergMeyerWasserman1985block}  introduced the first variants of the stochastic blockmodel  for directed networks, also log-linear in form,  which \cite{holland1983stochastic} extended by allowing latent block assignment. More recently,  \citet{YJFL-InferenceDirectedNtwkCovariates-JASA} discuss the asymptotic theory of the directed network model with covariates, a model whose log-linear interpretation is given in \cite{SteveAle:MLEloglin:AOS12}. 

The decades that followed \cite{FW81} saw a flurry of network modeling activity, starting with the early works of 
   \citet{WP1996logit}, \citet{PW1999logit}, and \citet{RPW1996logit} who introduced the more general class of ERGMs that included dyad dependent models,   
  building on Markov graphs  of  \citet{FS86}. 
  Since then,  the network modeling literature has  begun to address the  implementation of  goodness-of-fit and hypothesis testing algorithms for specific models, specific families of models, or specific tasks.  An early example includes the Conditional Uniform Graphs test \cite{AndersonBrighamS1999Tios}, which compares a graph statistic against an empirical distribution of the statistic obtained by randomly sampling uniformly from the space of all graphs with the same number of vertices and edges. More recently, \cite{Lei16} provides an asymptotic test for the basic variant stochastic blockmodel, along with a non-asymptotic version, while 
\cite{BanerjeeMa2017OptimalTests} offer a hypothesis test for the stochastic blockmodel vs.\ the Erd\"os-R\'enyi model, with optimality guarantees under certain degree growth conditions. 
For general ERGMs, the question of testing model goodness of fit has been generally acknowledged as difficult. 
It took until 2008 for the first comprehensive
 approach 
 to appear in \cite{Hunter} along with an {\tt R} package for  network models. 

\medskip 
Notable in the literature is the apparent divergence of focus between \emph{contingency table} models and \emph{network} models in practice. On the one hand, the network literature  generally embraces powerful and fast algorithms rooted in graph theory. Two illustrative examples, \cite{BlitzDiac10} and \cite{ZhangChenJASA13}, use  sequential importance sampling (SIS)  to sample graphs with a prescribed network statistic - 
the degree sequence in this case.  The approach, applicable to a small number of log-linear models, is computationally effective, and represents one of the two main approaches   to goodness-of-fit testing based on running a finite-sample algorithm; see also the package  \cite{networksis}. 
On the other end of the spectrum, the contingency table literature offers a series of exciting breakthroughs that apply directly  to the question  of goodness of fit. 
For \emph{all}  log-linear models for categorical data, 
the goodness-of-fit question
has been theoretically answered nearly three decades ago:   \citet{DS98} propose  a Monte Carlo Markov chain algorithm
to compute a  finite-sample approximation of an exact conditional test. 
The framework is guaranteed to work in all log-linear settings, though since it requires the computation of a Markov basis, methods for an  effective implementation  have remained elusive. 

 Our mission is to connect the two---graph-based and table-based approaches---and apply the combined theory to fitting a wide class of ERGMs.  We take inspiration from 
\cite{SteveAleMe-holland} and \cite{OHT}  (cf.\ \cite{PRF:09}, which appeared in the same volume as the latter),  who have adapted the Markov bases approach to network applications specifically, and who first demonstrated the feasibility of the method in practice despite apparent theoretical barriers to Markov bases, all of which we discuss below.

Section~\ref{sec:models} defines the class of log-linear models for networks in  Definition~\ref{defn:loglinearERGM}  and provides the correspondence to contingency tables explicitly. Definition~\ref{def:p1SBM} constructs a new log-linear ERGM, $p_1$-SBM,  encapsulating all `mix-and-match' model variants for `micro- and macro-analyses of sociometric relations' as suggested by \citet{FienbergMeyerWasserman1985block}. 
In Section~\ref{sec:GoFtesting}, we formulate a combinatorially modified Metropolis-Hastings algorithm for general log-linear ERGMs. The algorithm has two main parts: Algorithms~\ref{alg:MH} and~\ref{alg:NextMove}; the latter is a combinatorial formulation of an efficient Markov basis sampler for each model to which it is applied. We provide the   full theory and implementation for models discussed in Section~\ref{sec:models}, with proof of correctness for those models in Proposition~\ref{prop:symmetryForOurModels}. 
Theorem~\ref{thm:GoF} specifies exactly what needs to be implemented for any other log-linear ERGM in order for Algorithm~\ref{alg:MH} to produce, with probability $1$, the exact conditional $p$-value of model fit. 
We demonstrate the test on simulated and real data sets of medium to moderately large size. One of the data sets includes a large number of structural zeros in the model, rendering standard algebraic statistics methods for testing model fit inapplicable. The proof of Theorem~\ref{thm:GoF} discusses why our algorithm is still correct in this scenario, and how it extends the standard Markov basis chain. 
We close with a discussion of the larger methodological context of our contributions and an outline of some open problems that remain. 

Let us briefly address some common concerns with the algebraic statistics approach,  details of the solutions to which are in the technical sections that follow. 
While the method relying on Markov bases is theoretically sound, it can exhibit difficulties in implementation and practice, depending on the approach. For example,   \cite{ZhangChenJASA13} pointed to the computational complexity of obtaining a full Markov basis and   the limited results on mixing times. 
However, since the publication of \cite{ZhangChenJASA13}, there has been work to address both of these issues.
Indeed, dynamic approaches to generating Markov bases elements, such as  \cite{Dobra2012} and \cite{OHT}, remedy  the computational strain of computing full bases. More recently, \cite{GPS16} implement the exact conditional test for one variant of the $p_1$ random directed graph model  of \cite{HL81} with dyad-dependent reciprocation in a combinatorial fashion,  
avoiding the usual computational bottleneck by constructing  one Markov move at a time based on the current state of the chain, rather than precomputing the full basis. 
We take into account all of these developments within Algorithms~\ref{alg:MH} and~\ref{alg:NextMove}. 
A summary of the state of the art on Markov bases for general contingency tables appeared recently in \cite{MarkovBases25years}.

\section{Log-linear ERGMs} 
\label{sec:models}

Ever since 
\citet{Moreno1934}'s sociograms,  it is well known that network data are best represented as graphs of nodes---or vertices---and edges\footnote{In social sciences, nodes and edges are called \emph{actors} and  \emph{ties}; here we use the standard graph theory terminology, also more common in recent network science literature.}.
We denote the (fixed) number of nodes in the network by $n$, and the class of all graphs on $n$ nodes  by $\mathcal G_n$.  Depending on the application, one may restrict the set $\mathcal G_n$ to simple or multiple (valued), undirected or directed graphs, or a mixture of these, as desired for each model. 
An exponential family random graph model (ERGM) is a collection of probability distributions on the (possibly restricted) sample space $\mathcal G_n$ for which the probability of occurrence of each $g\in\mathcal G_n$ takes the 
form: 
\begin{equation}\label{eq:ERGM}
P_{\vec\theta}(G=g) = Z(\vec\theta) e^{{\vec\theta} \cdot t(g)} 
\end{equation}
 with  parameter vector $\vec\theta$, sufficient statistics $t(g)$, and normalizing constant  $Z(\vec\theta)$. 
 
Since the exponential family form is assumed, it is not unusual to specify models in the ERGM class by simply stating the vector of sufficient statistics $t(g)$, as we will do here.   Each entry of $t(g)$ is a network statistic, such as the edge count, degree of a given vertex,  number of edges in a given block of vertices, and so on. 


The general framework for the construction of an ERGM starts by first regarding each network \emph{dyad}---a pair of nodes---as a random variable, a step that implies `a stochastic framework with a fixed node set' \citep[\S2.1]{Robins2007}. 

%
\begin{defn}[Log-linear ERGMs]\label{defn:loglinearERGM}
Consider the ERGM  model family as in \eqref{eq:ERGM}. 
We call such a model a \emph{log-linear ERGM} if the sufficient statistic $t(g)$ is $(1)$ non-negative integer-valued in each coordinate and $(2)$ is additive over network dyad.
\end{defn}


Log-linear ERGMs belong to the class of dyad independent models. An ERGM is log-linear exactly when there exists a (redundant) contingency table representation  $u$ of the dyads of  $g$
 such that the sufficient statistic $t(g)$ corresponds to a set of table marginals $t(u)$.  An example of how one derives a dyadic contingency table from the network is included before Section~\ref{sec:knownmodels}, on page~\pageref{table example}.
 
The corresponding contingency table is the \emph{dyad classification table}: 
Two of the dimensions of the table index dyads (pairs of nodes), with additional dimensions indexing dyad configurations 
 and any possible block effects. 
 For dyad configurations, there can be a single category with $2$ levels---0 and 1---indicating presence or absence of an observed edge for undirected graphs, or there can be two categories of format $2\times2$ for directed graphs, as we will see below. 
 Block effects can also be categorized in $1$ or $2$ dimensions,  indexing a pair of blocks, the specifics depending on model specification. 
 Various  restrictions on the  sample space of graphs $\mathcal G_n$, such as disallowing multiple edges, etc., translate to sampling constraints on the space of tables $\mathcal U$ in the log-linear model. 
  For example, if the graph is assumed to have no multiple edges, then there are $0/1$ sampling constraints placed on the cells of the contingency table, while disallowing self-loops amounts to structural zeros on the  $n\times n$ diagonal. 
The connection to contingency tables is crucial 
   as it allows for the easy transfer of fitting and testing methods for contingency tables to graphs.  

For readers familiar with the algebraic statistics literature, the sufficient statistic being additive on the dyads means that 
the ERGM is \emph{a toric model on the dyad classification table}. Precisely, 
one embeds $\G_n \to (\Z_{\geq 0})^{\ell}$ by representing a graph $g$ as a vector whose entries are the dyads of $g$. Here the vector space dimension $\ell$ depends on the sample space; for example,   for the class of simple undirected graphs, for which the adjacency matrix is symmetric with all-zero diagonal entries, $ (\Z_{\geq 0})^{n \choose 2}$ could be sufficient. In Equation~\ref{eq:ERGM}, the map $t$ is then a map from an $\ell$-dimensional vector to another, say, $q$-dimensional vector of  sufficient statistics. Log-linearity then means the sufficient statistic map $t$ is a \emph{linear} map $t:  (\Z_{\geq 0})^{\ell} \to  (\Z_{\geq 0})^q$ from the  vectorized space of graphs  to the space of the minimal sufficient statistics of the model; this gives rise to the familiar design matrix of the model, the integer matrix with which we associate a toric variety in the standard way \citep{GeigerMeekSturmfels,DSS09}.

As an example, Definition~\ref{def:p1SBM} proposes the most general log-linear ERGM for simultaneously modeling node and group---or block---effects on dyad formations.  
\paragraph{Notation.}\label{dyadic notation} Since each dyad $(i,j)$ can be in 1 of 4 configurations: no link, directed from $i$ to $j$, directed from $j$ to $i$, or reciprocated, we say that a dyad $(i,j)$ can be in one of the following states: $(0,0)$, $(1,0)$, $(0,1)$, or $(1,1)$, respectively. 
Following \cite{HL81,FW81}, we denote $p_{ijkl}$ the probability of the dyad $(i,j)$ to be in state $(k,l)$: 
\[ 
	p_{ijkl} \equiv P\left( \mbox{dyad } (i,j) \mbox{ is in state } (k,l)\right).
\]

\begin{defn}[The $p_1$-SBM]  \label{def:p1SBM}
Suppose $n$ nodes are partitioned into $k$ blocks, and dyads in a directed network on the $n$ nodes are independent,  their formation  governed by the following three types of parameters, for $1\leq i,j\leq n$ and $1\leq r,s\leq K$: 
\begin{itemize}
\item $\alpha_i$ and $\beta_i$, propensity of the node $i$ to send and receive edges, respectively; 
\item  $\rho_{ij}$, edge reciprocation in the dyad $\{i,j\}$; 
\item  $\delta^{rs}$, expansiveness or density between blocks $r$ and $s$. 
\end{itemize}

The $p_1$-SBM model  postulates  the following dyad state probabilities: 
\begin{align}\label{eq:p1sbm}
	\log p_{ij00} &= \lambda_{ij}  \\
\nonumber	\log p_{ij10} &=  \lambda_{ij} + \alpha_i+\beta_j + \delta^{b(i)b(j)}\\
\nonumber		\log p_{ij01} &=  \lambda_{ij} +\alpha_j+\beta_i+ \delta^{b(j)b(i)} \\
\nonumber		\log p_{ij11}&= \lambda_{ij}+\alpha_i+\alpha_j +\beta_i+\beta_j + \delta^{b(i)b(j)}+\delta^{b(j)b(i)}+\rho_{ij}. 
\end{align}
\end{defn}

The  derivation of the likelihood function  of the network as a collection of independent dyads in exponential family form \eqref{eq:ERGM}  is now straightforward. 
The node-specific parameters resemble those in the $p_1$ model, while the block-specific parameters   $\delta^{rs}$ and $\delta^{sr}$ take place of the $\alpha_{rs}$ from the $\beta$-SBM; see next section for details. 

\paragraph{Model variants.} 
From this most general version of the $p_1$-SBM, one may specify several model variants.  The reciprocation parameter $\rho_{ij}$ can be set to be  constant across the network, i.e., $\rho_{ij}=\rho$, following Holland-Leinhardt, or, more interestingly, it can depend only on the block membership of the dyad, $\rho_{ij}=\rho^{b(i)b(j)}$, following Fienberg-Meyer-Wasserman. 
Then, 
  a sufficient statistics vector is 
\[
	t(G) = (\vec{d}^{\text{in}},\vec{d}^{\text{out}}, e_{1,1},e_{1,2},\dots,e_{K,K} ),  
\]
where $e_{k,l}$,  $1\leq k\leq l\leq K$, is the number of edges (edge density) 
 between and within blocks. 

Other variants of the model can be specified to allow the parameter interpretation to suit  the application at hand. For example, the block parameters can be set to ignore the direction by $\delta^{b(i)b(j)} = \delta^{b(j)b(i)}$; the corresponding statistic counts the total number of edges between a pair of blocks, regardless of direction. Or, the reciprocation parameter can be defined to be  block-specific, measuring the tendency of each block of nodes to reciprocate edges,  by setting  $\rho^{b(i)b(j)}=\rho^{b(i)}+\rho^{b(j)}$. In this case, 
 the sufficient statistic vector would include the number of reciprocated edges incident to each block. 
Setting the node-specific parameters $\alpha_i$ and $\beta_j$ to zero recovers the general blockmodel \cite{FienbergMeyerWasserman1985block}, while setting the block-specific parameters $\delta^{rs}$ to zero recovers the  $p_1$-model \cite{FW81}; therefore this is a generalization of both.

\paragraph{Log-linearity.}
\begin{wrapfigure}{r}{0.35\textwidth}
\vspace{-8mm}
\centering
\begin{tikzpicture}[scale=1, every node/.style={minimum size=5mm, text centered}]
   \foreach \i/\color in {1/blue, 2/blue} {
       \node[draw, circle, preaction={fill=\color, fill opacity=0.1}, draw=\color, inner sep=0pt] (n\i) at ({(-\i)*360/4+180}:1.7) {{\i}};
   }
   \foreach \i/\color in {3/red, 4/red} {
       \node[draw, circle, preaction={fill=\color, fill opacity=0.1}, draw=\color, rectangle, inner sep=0pt] (n\i) at ({(-\i)*360/4+180}:1.7) {{\i}};
   }
\tikzset{edge/.style = {->,> = stealth'}}
   
	\draw[edge] (n1) to (n2);
	\draw[edge] (n2) to (n1);
	\draw[edge] (n1) to (n3);
	\draw[edge] (n1) to (n4);
	\draw[red, dashed] [edge] (n4) to (n2);
\end{tikzpicture}
\caption{A $4$-node network with two blocks: the blue block consists of circular nodes  $1$ and $2$, and the red block consists of rectangular nodes $3$ and $4$.} 
\label{fig:example-graph}
\vspace{-3mm}
\end{wrapfigure}
Following the format of the dyad classification table described below Definition~\ref{defn:loglinearERGM} on page~\pageref{defn:loglinearERGM}, the $p_1$-SBM is equivalent to a log-linear model on a contingency table which is a redundant representation of the adjacency matrix of the graph. 
Let us construct this table explicitly for one of the variants in this model family. 
\label{table example}
We illustrate this with a small example, a $2$-block network on four nodes in Figure~\ref{fig:example-graph} with four edges, one of which -- the edge between nodes 1 and 2 -- is reciprocated, or bidirected. 
The corresponding contingency table will be of the format $4\times4\times2\times2\times3$.  Following the convention for dyadic notation from Page~\pageref{dyadic notation} (see ``Notation"), the $2\times2$ part encodes the state of each dyad in one of the following ways: 
\begin{table}[!h]
\centering
Table configuration at $(u,v,\cdot,\cdot,b)$ 
\qquad \quad 
\scalebox{0.3}{
	\begin{tabular}{|l|l|}
	\hline
	\cellcolor{black} &  \\ \hline
	\phantom{x} & \phantom{x} \\ \hline
	\end{tabular}
	}
\qquad
\scalebox{0.3}{
	\begin{tabular}{|l|l|}
	\hline
	& \cellcolor{black}  \\ \hline
	\phantom{x} & \phantom{x} \\ \hline
	\end{tabular}
	}
\qquad
\scalebox{0.3}{
	\begin{tabular}{|l|l|}
	\hline
	\phantom{x} & \phantom{x} \\ \hline
	\cellcolor{black} &  \\ \hline
	\end{tabular}
	}
\qquad
\scalebox{0.3}{
	\begin{tabular}{|l|l|}
	\hline
	\phantom{x} & \phantom{x} \\ \hline
	& \cellcolor{black}  \\ \hline
	\end{tabular}
	}.

State of the dyad $u,v$ in block slice $b$:  \quad (0,0) \quad (0,1)  \quad  (1,0)  \quad (1,1) 
\end{table}

\noindent 
For simple graphs, all non-zero entries of the $2\times 2$ part of the table  are set to $1$, as a sampling constraint. In addition, since each dyad can be observed in only one state at a time, only one of the entries of the $2\times2$ part of the table will be nonzero. The nonzero cell is shaded black in the figure.

In the table, each edge is represented twice: once as a state of  the ordered dyad $(u,v)$ and once  as $(v,u)$. 
The last dimension of the table, $3$, encodes  the block pair of the dyad. Since there are two blocks, blue and red, there are three block-pairs: blue/blue, red/red, and blue/red. The following are then the slices in which each dyad can have a non-zero entry: 
\begin{table}[!h]
\small
\begin{tabular}{r |c |c |c}
block-pair slice & dyad & state & $2\times2$ slice 
\\
\hline
blue & 12 & (1,1)  &\scalebox{0.3}{
	\begin{tabular}{|l|l|}
	\hline
	\phantom{x} & \phantom{x} \\ \hline
	& \cellcolor{black}  \\ \hline
	\end{tabular}
	} \\  
blue/red & 13 & (0,1)&  \scalebox{0.3}{
	\begin{tabular}{|l|l|}
	\hline
	& \cellcolor{black}  \\ \hline
	\phantom{x} & \phantom{x} \\ \hline
	\end{tabular}
	}\\
blue/red  & 14 & (0,1)& \scalebox{0.3}{
	\begin{tabular}{|l|l|}
	\hline
	& \cellcolor{black}  \\ \hline
	\phantom{x} & \phantom{x} \\ \hline
	\end{tabular}
	}\\
blue/red  & 23&(0,0)& \scalebox{0.3}{
	\begin{tabular}{|l|l|}
	\hline
	\cellcolor{black} &  \\ \hline
	\phantom{x} & \phantom{x} \\ \hline
	\end{tabular}
	}
\\
blue/red &24&
		{\color{red}(1,0)}& 
  {\color{red}\scalebox{0.3}{
	\begin{tabular}{|l|l|}
	\hline
	\phantom{x} & \phantom{x} \\ \hline
	\cellcolor{red} &  \\ \hline
	\end{tabular}
	}
}\\
red&34&(0,0)& \scalebox{0.3}{
	\begin{tabular}{|l|l|}
	\hline
	\cellcolor{black} &  \\ \hline
	\phantom{x} & \phantom{x} \\ \hline
	\end{tabular}
	}
\end{tabular}
\qquad 
\begin{tabular}{r |c |c |c}
block-pair slice & dyad & state & $2\times2$ slice 
\\
\hline
blue & 21 & (1,1)  &\scalebox{0.3}{
	\begin{tabular}{|l|l|}
	\hline
	\phantom{x} & \phantom{x} \\ \hline
	& \cellcolor{black}  \\ \hline
	\end{tabular}
	} \\  
blue/red  & 31 & (1,0)  &\scalebox{0.3}{
	\begin{tabular}{|l|l|}
	\hline
	\phantom{x} & \phantom{x} \\ \hline
	\cellcolor{black} &  \\ \hline
	\end{tabular}
	}
\\
blue/red  & 41 & (1,0) &\scalebox{0.3}{
	\begin{tabular}{|l|l|}
	\hline
	\phantom{x} & \phantom{x} \\ \hline
	\cellcolor{black} &  \\ \hline
	\end{tabular}
	}
\\
blue/red  & 32&(0,0) &\scalebox{0.3}{
	\begin{tabular}{|l|l|}
	\hline
	\cellcolor{black} &  \\ \hline
	\phantom{x} & \phantom{x} \\ \hline
	\end{tabular}
	}
\\
blue/red &42&
		{\color{red}(0,1)} & 
 {\color{red}\scalebox{0.3}{
	\begin{tabular}{|l|l|}
	\hline
	& \cellcolor{red}  \\ \hline
	\phantom{x} & \phantom{x} \\ \hline
	\end{tabular}
	}}\\
red&43&(0,0) &\scalebox{0.3}{
	\begin{tabular}{|l|l|}
	\hline
	\cellcolor{black} &  \\ \hline
	\phantom{x} & \phantom{x} \\ \hline
	\end{tabular}
	}

\end{tabular}
\end{table}

\noindent 
The red entries in the table represent the red dashed edge in the graph. 

Next, we will fill in the $4\times 4$ part of the contingency table. Rows and columns here are indexed by the $4$ nodes. The full contingency table is listed in Table~\ref{full contingency table}. 
\begin{table}[!h]
\footnotesize
The 
\scalebox{0.3}{
\begin{tabular}{|l|l|}
\hline
\cellcolor{black} &  \\ \hline
\phantom{x} & \phantom{x} \\ \hline
\end{tabular}
}
slice consists of  $4\times4\times3$ table:   
\quad   	
	\begin{tabular}{|c|c|c|c|}
	\multicolumn{4}{c}{Edges within }\\
	\multicolumn{4}{c}{{\color{blue}blue} block}  \\ \hline
	0 & 0 & 0 & 0 \\ \hline
	0 & 0 & 0 & 0 \\ \hline
	0 & 0 & 0 & 0 \\ \hline
	0 & 0 & 0 & 0 \\ \hline
	\end{tabular}
\quad 
	\begin{tabular}{|c|c|c|c|}
	\multicolumn{4}{c}{Edges within }\\
	\multicolumn{4}{c}{{\color{red}red} block}  \\ \hline
	0 & 0 & 0 & 0 \\ \hline
	0 & 0 & 0 & 0 \\ \hline
	0 & 0 & 0 & 1 \\ \hline
	0 & 0 & 1 & 0 \\ \hline
	\end{tabular}
\quad 
	\begin{tabular}{|c|c|c|c|}
	\multicolumn{4}{c}{Edges {\color{purple}between} }\\
	\multicolumn{4}{c}{{\color{red}red} and {\color{blue}blue}}  \\ \hline
	0 & 0 & 0 & 0 \\ \hline
	0 & 0 & 1& {\color{red}\bf 0} \\ \hline
	0 & 1 & 0 & 0 \\ \hline
	0 & {\color{red}\bf 0} & 0 & 0 \\ \hline
	\end{tabular}

The 
\scalebox{0.3}{
	\begin{tabular}{|l|l|}
	\hline
	& \cellcolor{black}  \\ \hline
	\phantom{x} & \phantom{x} \\ \hline
	\end{tabular}
	}
slice consists of  $4\times4\times3$ table:   
\quad   	
	\begin{tabular}{|c|c|c|c|}
	\multicolumn{4}{c}{{\color{blue}blue}}  \\ \hline
	0 & 0 & 0 & 0 \\ \hline
	0 & 0 & 0 & 0 \\ \hline
	0 & 0 & 0 & 0 \\ \hline
	0 & 0 & 0 & 0 \\ \hline
	\end{tabular}
\quad 
	\begin{tabular}{|c|c|c|c|}
	\multicolumn{4}{c}{{\color{red}red}}  \\ \hline
	0 & 0 & 0 & 0 \\ \hline
	0 & 0 & 0 & 0 \\ \hline
	0 & 0 & 0 & 1 \\ \hline
	0 & 0 & 0 & 0 \\ \hline
	\end{tabular}
\quad 
	\begin{tabular}{|c|c|c|c|}
	\multicolumn{4}{c}{{\color{purple}between}}  \\ \hline
	0 & 0 & 1 & 1 \\ \hline
	0 & 0 & 0 & 0 \\ \hline
	0 & 0 & 0 & 0 \\ \hline
	0 & {\color{red}\bf 1}  & 0 & 0 \\ \hline
	\end{tabular}

The 
\scalebox{0.3}{
	\begin{tabular}{|l|l|}
	\hline
	\phantom{x} & \phantom{x} \\ \hline
	\cellcolor{black} &  \\ \hline
	\end{tabular}
	}
slice consists of  $4\times4\times3$ table:   
\quad   	
	\begin{tabular}{|c|c|c|c|}
	\multicolumn{4}{c}{{\color{blue}blue}}  \\ \hline
	0 & 0 & 0 & 0 \\ \hline
	0 & 0 & 0 & 0 \\ \hline
	0 & 0 & 0 & 0 \\ \hline
	0 & 0 & 0 & 0 \\ \hline
	\end{tabular}
\quad 
	\begin{tabular}{|c|c|c|c|}
	\multicolumn{4}{c}{{\color{red}red}}  \\ \hline
	0 & 0 & 0 & 0 \\ \hline
	0 & 0 & 0 & 0 \\ \hline
	0 & 0 & 0 & 0 \\ \hline
	0 & 0 & 1 & 0 \\ \hline
	\end{tabular}
\quad 
	\begin{tabular}{|c|c|c|c|}
	\multicolumn{4}{c}{{\color{purple}between}}  \\ \hline
	0 & 0 & 0 & 0 \\ \hline
	0 & 0 & 0 & {\color{red}\bf 1}\\ \hline
	1 & 0 & 0 & {\color{red}\bf 1} \\ \hline
	1 & 0  & 0 & 0 \\ \hline
	\end{tabular}
	
The 
\scalebox{0.3}{
	\begin{tabular}{|l|l|}
	\hline
	\phantom{x} & \phantom{x} \\ \hline
	& \cellcolor{black}  \\ \hline
	\end{tabular}
	}
slice consists of  $4\times4\times3$ table:   
\quad   	
	\begin{tabular}{|c|c|c|c|}
	\multicolumn{4}{c}{{\color{blue}blue}}  \\ \hline
	0 & 1 & 0 & 0 \\ \hline
	1 & 0 & 0 & 0 \\ \hline
	0 & 0 & 0 & 0 \\ \hline
	0 & 0 & 0 & 0 \\ \hline
	\end{tabular}
\quad 
	\begin{tabular}{|c|c|c|c|}
	\multicolumn{4}{c}{{\color{red}red}}  \\ \hline
	0 & 0 & 0 & 0 \\ \hline
	0 & 0 & 0 & 0 \\ \hline
	0 & 0 & 0 & 0 \\ \hline
	0 & 0 & 0 & 0 \\ \hline
	\end{tabular}
\quad 
	\begin{tabular}{|c|c|c|c|}
	\multicolumn{4}{c}{{\color{purple}between}}  \\ \hline
	0 & 0 & 0 & 0 \\ \hline
	0 & 0 & 0 & 0 \\ \hline
	0 & 0 & 0 & 0 \\ \hline
	0 & 0  & 0 & 0 \\ \hline
	\end{tabular}
\caption{The $4\times4\times2\times2\times3$ dyad classification table for the graph in Figure~\ref{fig:example-graph}. The bold red entries indicate the cells that change when the dashed red edge {\color{red}$3\to 4$} is present in the graph.}\label{full contingency table}
\end{table} 
For completeness, let us consider a table margin, for example, the $[34]$-margin of this table would be: 
	\begin{tabular}{|l|l|}
	\hline
	4 & 4\\ \hline
	4 & 2   \\ \hline
	\end{tabular}. 
The top left entry indicates there are 4 \emph{ordered} dyads in the state $(0,0)$, encoding  the two missing edges: $(2,3),(3,2),(3,4),(4,3)$. 

In general, when $\rho_{ij}=\rho^{b(i)b(j)}$ is block-constant,
we get an \mbox{$n\times n\times 2\times 2 \times (K+{K\choose 2})$} table.
 The first two table dimensions index nodes, the second two the 4 dyad configurations for each ordered pair of nodes, and the last one the pair of blocks to which a dyad belongs. 
The block-constant $p_1$-SBM's sufficient statistics are then encoded by the following table marginals:  
$[125]$, 
$[345]$, 
$[134]$ and $[234]$. 
As is standard in categorical data analysis, $[ijk]$ stands for the marginal of the table $u$ computed with  respect to the $i$-th, the $j$-th, and the $k$-th variables:  $\sum_{l,m}u_{ijklm}$.  
For this model, the margin $[125]$  fixes the number of edges observed per dyad. 
The margin $[345]$ records the edge count between each pair of blocks.
The last two, $[134]$ and $[234]$, represent the in-degrees and out-degrees of the nodes in the graph, respectively. 
 The table representation utilizes dyadic independence and the fact that ERGMs are models on the level of dyads as random variables. The following section contains many more specific examples.  

\paragraph{A note on dyadic independence.} 
 ERGMs  assuming independent dyads represent a broad and expressive class of models, including degree-based models such as the $\beta$ and $p_1$ models, and edge-count-based models such as the stochastic blockmodels.  They also include degree-based extensions of blockmodels, as well as extensions that include covariates as discussed by \citet{YJFL-InferenceDirectedNtwkCovariates-JASA}.  
 In the context of \citet{Shalizi2013}, {dyadic independence} in an log-linear ERGM is equivalent to the sufficient statistic $t(G)$ adding up over the dyads in the network. 
While \emph{dyad-dependent} ERGMs that include $k$-stars or triangle counts in the sufficient statistics have further expressive power, some care should be taken in estimating them; model degeneracy and asymptotics are further discussed in other work including \cite{Chatterjee2013}, \cite{Handcock2003}, \cite[\S 1.1]{YJFL-InferenceDirectedNtwkCovariates-JASA}, and \cite{SchweinbergerEtAl2020}.

\subsection{Some popular ERGMs and  their log-linear  representations}\label{sec:knownmodels}

\paragraph{The $\beta$-model.} This is the simplest degree-based model for undirected graphs and it assumes independent dyads. The name for the model was coined by \citet{CDS11}, who restrict the sample space $\mathcal G_n$ to simple graphs, that is, graphs without multiple edges or self-loops. \citet{RPF:11} extend the model for multiple edge counts. 
The sufficient statistic for the $\beta$-model for undirected graphs is the degree sequence $t(G)= (d_1, \dots, d_n)$, with $d_i$ counting the number of edges incident to vertex $i$.  Associated to each vertex $i$  is a parameter $\beta_i$ that controls for the propensity of  $i$ to form edges.  In exponential family form, the model is
$$P_{\beta_1, \beta_2,\ldots, \beta_n}(G=g) = Z(\beta_1, \beta_2,\ldots, \beta_n) e^{d_1\cdot \beta_1 + \ldots d_n \cdot \beta_n}, \quad\quad \beta_i\in\R.$$

Dyadic independence leads to an alternative specification of the model, 
using the probabilities 
$p_{i,j}$ of each dyad $ij$ being connected in $g$   or, equivalently, their log-odds:
$$p_{i,j}= \frac{e^{\beta_i+\beta_j}}{1+e^{\beta_i+\beta_j}} \quad\quad \mbox{equiv.,} \quad\quad    \log \left( \frac{p_{i,j}}{1-p_{i,j}} \right) = \beta_i + \beta_j, \quad\quad i,j\in\{1,2, \ldots n\}.$$ 
We will express the  contingency table representation of  the $\beta$-model  as a special case of $p_1$. 

\paragraph{The $p_1$-model.} 
The $p_1$-model, introduced by \citet{HL81},  is a direct generalization of the above to directed graphs.  
Using the same dyad-state notation introduced for Definition~\ref{def:p1SBM} on page~\pageref{def:p1SBM}, the dyadic probabilities in the $p_1$-model are  specified as:
\begin{align}\label{eq:p1}
	\log p_{ij00} &= \lambda_{ij},  \\
\nonumber	\log p_{ij10} &=  \lambda_{ij} + \alpha_i+\beta_j+\delta,\\
\nonumber		\log p_{ij01} &=  \lambda_{ij} +\alpha_j+\beta_i+\delta,\\
\nonumber		\log p_{ij11}&= \lambda_{ij}+\alpha_i+\alpha_j+\beta_j+\beta_j+2\delta+\rho_{ij}. 
\end{align}
The model assumes  dyadic independence. Of course, expressing the likelihood function of the $p_1$ ERGM is a straightforward exercise.  
  \citet{FW81} show the model's equivalence to a log-linear model on an $n\times n\times 2\times 2$ contingency table (see paragraph following Definition~\ref{defn:loglinearERGM}). 
  
There are three types of parameters in $p_1$. 
Node-specific parameters $\alpha_i$  and $\beta_i$ record the propensities of  the node $i$  sends and receives links, respectively.  
The  parameter $\rho_{ij}$ encodes the edge reciprocation  effect, allowing for three main model variants:  zero reciprocation, constant reciprocation, and dyad-specific reciprocation (to which Fienberg and Wasserman referred as `differential reciprocity').     
Additional parameters $\delta$ and $\lambda_{ij}$ are normalizers.  The density parameter $\delta$, as \cite{HL81} put it, is ``similar to the notion of gross expansiveness in the affective sociometric context'', and is also called the ``choice parameter'' in \cite{FW81}.  
The ${n\choose 2}$ dyadic effects, $\lambda_{ij}$,  are normalizing constants that ensure the four dyad probabilities add to $1$ for each dyad. These parameters will, of course, add corresponding sufficient statistics to the vector $\theta$, however, they are not minimal: for example, the sufficient statistics corresponding to $\delta$ is the total number of edges in the graph, which can be computed from node degrees. 

 From~\eqref{eq:p1}, one easily recovers both the sufficient statistics 
  and equivalent contingency table marginals \citep[Equations (5), (6), and (8)]{FW81}, which we spell out for the sake of  completeness.  
In the contingency table setting, the three variants of the model 
---  zero ($\rho_{ij}=0$), constant ($\rho_{ij}=\rho$), and dyad-specific ($\rho_{ij}=\rho_i+\rho_j$) reciprocation --- 
 correspond to log-linear models specified by following marginals: $[12][13][14][23][24]$, $[12][13][14][23][24][34]$, and $[12][134][234]$, 
 respectively, of the 4-way table $u$. 
   By considering only reciprocated edges, a simple parameter substitution recovers the $\beta$-model as a submodel of $p_1$ with a similar contingency table representation: $n\times n\times 2$ table with the following marginals as sufficient statistics: $[13],[23],[12]$.

 The sufficient statistic $t(G)$  for the $p_1$-model with \emph{zero reciprocation} consists of the out-degree sequence and in-degree sequence, also known as the bidegree sequence in graph theory: 
\[t(G)=(\vec{d}^{\text{in}},\vec{d}^{\text{out}})=(d_1^{in}, \ldots d_n^{in}, d_1^{out}, \ldots, d_n^{out}).\]
The corresponding model parameter vector 
is $\theta=(\alpha_1, \ldots, \alpha_n, \beta_1, \ldots, \beta_n)$.
  The sufficient statistic for the $p_1$-model with \emph{constant reciprocation} consists of the bidegree sequence and the reciprocated edge count $m$: 
\[t(G)=(d_1^{in}, \ldots d_n^{in}, d_1^{out}, \ldots, d_n^{out}, m) = 
(\vec{d}^{\text{in}},\vec{d}^{\text{out}}, m)\] 
and the model parameter vector is specified as $\theta=(\alpha_1, \ldots, \alpha_n, \beta_1, \ldots, \beta_n, \rho)$.  The additional parameter $\rho$ is the reciprocation parameter identifying the overall propensity of nodes to reciprocate links, which is assumed to be constant across the entire network. 
The sufficient statistic for the $p_1$-model with \emph{dyad-specific reciprocation} consists of the reciprocated degree sequence and the number of reciprocated edges $m_i$ incident to each vertex: 
\[t(G)=(\vec{d}^{\text{in}},\vec{d}^{\text{out}}, \vec{m})=
(d_1^{in}, \ldots d_n^{in}, d_1^{out}, \ldots, d_n^{out}, m_1, \ldots, m_n)\] 
with model parameter vector $\theta=(\alpha_1, \ldots, \alpha_n, \beta_1, \ldots, \beta_n, \rho_1, \ldots, \rho_n)$.
 Here, each parameter $\rho_i$ encodes the rate at which  node $i$ is likely to reciprocate links,  
  thus  allowing one to record different reciprocation effects for each dyad. 

\paragraph{The blockmodel.} 
Blockmodels are natural models for network data in which nodes belong to groups, or blocks, according to nodal attributes. 
Models for relational data where node attributes are \emph{latent}  address a certain set of applications whose aim is different in nature  from the applications we address here; see \citet{NowSniJASA01} for an overview. 
 The general form of a dyadically-independent stochastic blockmodel was introduced in \cite{FienbergWasserman1981categorical}  and  discussed, along with its latent version,  in the foundational papers \cite{holland1983stochastic} and \cite{FienbergMeyerWasserman1985block}. 
 The latter, in fact, showed that the non-latent versions of these models are log-linear in form. 
   
As in the $p_1$-model above, we use $p_{ijkl}$ to denote the probability of the dyad $(i,j)$ to be in state $(k,l)$ where $(k,l) \in \{0,1\}^2$. Note that if the network is undirected, the model simply collapses to having only two dyadic states: (0,0) and (1,1).  \citet{FienbergMeyerWasserman1985block}'s  general class of models with nodal attributes that partition nodes into blocks specifies dyad probabilities as follows: 

\begin{align}\label{eq:FMWblockmodel}
	\log p_{ij00} &= \lambda_{ij}  \\
\nonumber	\log p_{ij10} &=  \lambda_{ij} + \delta^{b(i)b(j)}\\
\nonumber		\log p_{ij01} &=  \lambda_{ij} + \delta^{b(j)b(i)}\\
\nonumber		\log p_{ij11}&= \lambda_{ij}+\delta^{b(i)b(j)}+\delta^{b(j)b(i)}+\rho^{b(i)b(j)}, 
\end{align}
where   each node in the graph belongs to one of $K$ blocks, $B_1,\dots,B_K$, and $b(i)$ denotes the  (known) block assignment of vertex $i$. 
 \citet{FienbergWasserman1981categorical} refer to parameters $\delta^{rs}$, $1\leq r,s,\leq K$, as \emph{choice effects}, and the $\rho^{rs}$ as reciprocity effects. However, unlike in the standard $p_1$-model, the reciprocity effects are on the level of \emph{blocks} rather than  \emph{nodes}. 

There are now various special cases: one can assume that the reciprocity parameters are all equal for any pair of blocks $B_r$ and $B_s$, resulting in only one parameter $\rho$ constant across the network; or assume an additive structure akin to what Holland and Leinhardt did for $p_1$; or something entirely different. 
For example,  choosing $\delta^{rs}=\delta+\alpha^r+\beta^s$ and $\rho^{rs}=\rho$, as in \citet[Equation (2.10)]{FienbergMeyerWasserman1985block}, provides what looks like a block-version of Holland and Leinhardt's $p_1$ -- in fact, it's precisely the $p_1$ on the blocks rather than on the nodes. This model was also suggested in the comment by \citet{BreigerHLcommentJASA81}. 
The sufficient statistics are the block in-degrees and the block out-degrees, rather than node in- and out-degrees, and the total number $m$ of reciprocated edges in the network:  
\[
	t(G) = ({\bf d}_{B}^{in},{\bf d}_B^{out},m) := \left( d_{B_1}^{in},\dots,d_{B_k}^{in}, d_{B_1}^{out},\dots,d_{B_k}^{out} ,m\right),
\]
where  the in-degree of block $B_j$ is $d_{B_j}^{in}=\sum_{i\in B_j} d_i^{in}$, the sum of degrees of all nodes in the block, and the out-degree is defined similarly. 

While the block models above do not use any node-specific parameters,  the authors note that ``there are more elaborate subgroup models [...] that combine individual actor with subgroup parameters'', although they do not discuss them explicitly. 
We discuss them next. 

\paragraph{The $\beta$-SBM.} 
The \emph{$\beta$-SBM} combines the basic undirected blockmodel above  with the $\beta$-model, thus incorporating the node effects into the modeling framework. Its exponential family form was formally introduced in \cite{Xiaolin2015thesis}. 
This is simply  the exponential family variant  of the famous \emph{degree-corrected stochastic blockmodel} (\cite{karrer2011stochastic}, see also \cite{ChoiWolfeAiroldiNIPS2011}) that has seen numerous applications. Instead of inferring block structure from the relational data akin to some popular community detection algorithms that relate to the degree-corrected SBM, we use background information on  the nodes -- node attributes -- to place them into natural blocks instead (cf. \cite{FienbergWasserman1981categorical}). 
From the point of view of neuronal data (e.g., Section~\ref{sec:applications}), this consideration is quite natural, as node attributes for groupings are  given by the data. 

 The $\beta$-SBM  postulates that each node in the graph belongs to one of $K$ blocks, $B_1,\dots,B_K$. Denoting by $b(i)$ the (known) block assignment of vertex $i$, the $\beta$-SBM gives the following log-odds for the probability $p_{ij}$ of each dyad $ij$ being connected: 
\begin{equation} 
\label{bSBM}
\log \left( \frac{p_{ij}}{1 - p_{ij}} \right) = \beta_i + \beta_j + \alpha_{b(i),b(j)}. 
\end{equation}
The node-specific parameters $\beta_i$ play the same role as in the $\beta$-model, while the block-specific parameters $\alpha_{b(i),b(j)}$ encode the propensities of blocks $i$ and $j$ being connected and represent the undirected version of $\delta^{b(i)b(j)}$ from Equation~\eqref{eq:FMWblockmodel}.  
The ${n\choose 2}$ equations \eqref{bSBM} define an exponential family model in which the minimal sufficient statistics are the degree sequence of the nodes and the number of edges between and within blocks, i.e., $|E_{k,l}|$ for $1\leq k\leq l\leq K$: 
\[
	t(G)=(d_1,\dots,d_n, |E_{1,1}|,|E_{1,2}|,\dots,|E_{K,K}|).
\]	

 The reader should note that restrictions should be placed on the parameter space to ensure identifiability of natural parameters in all of these models; for example, the $p_1$-model assumes $\sum\alpha_i=\sum\beta_i=0$. For our purposes of testing model fit, we work with the mean value parametrization, so we do not concern ourselves with the natural parameter restrictions. 

 Using the dyad classification table representation as before, we arrive at the log-linear model on contingency tables equivalent to the $\beta$-blockmodel: the table dimensions are $n\times n\times 2 \times (K+{K\choose 2}) $ and the marginals are
$[13], [23], [34], [124]$. 

\medskip

The $\beta$-SBM  is an extension of both the usual $\beta$-model and the simple blockmodel, as it combines both individual parameters  $\beta_i$, $1\leq i\leq n$, and the group interaction parameters $\alpha_{rs}$, $1\leq r,s\leq K$. 
There is a natural extension of this model that allows for directed links, so that the symmetry $\mbox{logit}(p_{ij})= \mbox{logit}(p_{ji})$ no longer holds. 
There are several ways to similarly extend the $p_1$-model by incorporating additional group-interaction parameters. In fact, this  is the direct motivation behind our $p_1$-SBM general model class.

\section{Goodness-of-fit testing}
\label{sec:GoFtesting}
Goodness-of-fit tests can be constructed in a few general ways, among which we choose the  conditional test. This classical approach applies in the same way  to all log-linear models: it computes the $p$-value of model fit based on the conditional distribution given sufficient statistics. 
In algebraic statistics, the support of this conditional distribution is called a \emph{fiber} of the observed graph $g_0$ with respect to  the log-linear model with sufficient statistics $t$:  
\begin{equation}\label{eq:fiber}
\F_t(g_0) = \{ g \in \mathcal G_n \ : \ t(g) = t(g_0) \}. 
\end{equation}
The set $\F_t(g_0)$ in Equation~\eqref{eq:fiber} is combinatorial in nature, as it is the set of all graphs with the same network statistics (e.g., edge count between/within blocks, degree sequence, etc.) as the observed graph $g_0$.  We refer to this reference set as the \emph{fiber of $g_0$ with respect to $t$}.  

The goodness-of-fit (GoF hereafter) problem requires the specification of a GoF statistic, a measure of how well the model fits the data. As per \cite{HL81}, who discuss some GoF statistics proposed in the decade prior to their work, a summary of which appears in Fienberg's subsequent work including  \citet[Section 2.1]{PRF:09}, the chi-square statistic is one reasonable choice. Namely, for the models considered in this paper, $\chi^2$ is not constant on non-trivial fibers (cf.~Subsection~\ref{sec:GoFknownmodels}), and as such can provide a theoretically valid $p$-value. We say `can' and not `does' here because of an obvious delicacy: classically, one declares model a poor fit if the observed value of a GoF statistic is statistically large, when compared to asymptotic approximations of the distribution of the GoF on the fiber. In networks, however, and in sparse contingency tables more generally, the use of such asymptotics is questionable.  
Thus, to compute the conditional $p$-value for GoF of the model, one instead computes the exact conditional distribution of the GoF on the fiber  and uses that as a reference instead. 

The combinatorial explosion of fiber size poses a practical challenge in implementing this test, requiring sampling the fiber  $\F_t(g_0)$ to  approximate the exact conditional $p$-value: 
$$Prob\{\chi^2(G) \geq \chi^2(g_0) \ | \ t(G) = t(g_0)\}.$$
The sample from the fiber is used to estimate the sampling distribution of a GoF statistic of choice, the observed value of which is then compared to the reference distribution of  sampled  values.  

Approximating the exact conditional test by sampling model fibers has been popular in contingency table analysis and particularly in the algebraic statistics literature. For any log-linear model, a Metropolis-Hastings algorithm can be used, as long as one has access to a way to sample from the fiber~\eqref{eq:fiber} of the observed graph.
A \emph{Markov basis} for a log-linear model is defined as  a set of moves guaranteed to connect \emph{each} fiber, meaning that the set $\F_t(g)$ is connected by elements in this basis for any observable $g$. 
 \citet{DS98} translated the notion of a Markov basis to an algebraic object, namely, a basis of a polynomial ideal. 
Thus a Markov basis exists and is finite for any log-linear model, precisely because of the algebra connection: 
\emph{For any log-linear model, a Markov basis exists and is finite by the Hilbert basis theorem in commutative algebra.} Specifically, it can be obtained through a generating set of the polynomial ideal defining the model.  
\noindent  Such a basis is a critical input to the GoF algorithm and can, in principle, be computed for any model.

In practice, there are naturally arising concerns about using  Markov bases, outlined in  \cite{F-review}, mainly revolving around  scalability  and difficulty of computation due to sampling constraints. Based on this, \citet{ZhangChenJASA13} erroneously concluded that that ``it is unclear whether proposals in this [algebraic statistics] literature are in fact reaching all possible tables associated with the distribution", whereas  the algebraic statistics literature      provides \emph{proof} that the chains \emph{are}, in fact, reaching all possible tables. Specifically, the connection to algebra established in \cite{DS98} implies explicitly that Markov chains built on Markov bases  
 are irreducible for any log-linear model, 
     and \citet[Proposition 2.1]{HT:10}  prove that in case of simple graphs, 
     chains built on Graver bases  are irreducible. Graver bases are much larger collections of moves than a Markov bases, but still finite and in principle computable using toric algebra. For more on Markov bases and Graver bases in algebraic statistics see \cite{Aoki2012}. 
     Note that when the sample space $\G_n$ consists of {simple} graphs, $g_0$ can be regarded as a $0/1$ contingency table and this conditional distribution is uniform; so one can use alternatives to Markov/Graver bases by importing fast graph-theoretic algorithms, when such algorithms are available. 
      For example,  efficient uniform sampling of graphs with a prescribed degree sequence remains an active area of research, recent progress of which can be traced through  \cite{BlitzDiac10}, \cite{Erdos2015},  and \cite{Sampling2012}.

\medskip 

In Algorithm~\ref{alg:MH}, whose correctness is proved in Theorem~\ref{thm:GoF},  we propose a general method for testing goodness of fit of a log-linear ERGM. 
The method avoids two  crucial issues of the usual Markov basis chain, both of which are prominent in ERGMs. The first is stated already in \cite{DS98}:   reliance on precomputing a (minimal) Markov basis; see also  the last paragraph of \cite{Dobra2012}. In contrast, the combinatorial subroutine Algorithm~\ref{alg:NextMove}
constructs (not necessarily minimal) Markov moves one at a time based on the current state of the chain. 
Second, a minimal Markov basis need not connect any fiber in the presence of sampling constraints; in contrast, our method Algorithm~\ref{alg:NextMove} incorporates such constraints in the dynamic construction of moves, guaranteeing connectivity of any  fiber possibly constrained by the structure of the sample space $\mathcal G_n$, which we discussed at the beginning of Section~\ref{sec:models}. 

In Step~\ref{step:selectSubtable} of Algorithm~\ref{alg:MH}, by a `random subtable' we mean a subtable of the dyad classification table such that it represents a random subgraph of the observed graph. The algorithm for selecting a random subgraph is specified by the user; for example, an Erd\H{o}s-R\'enyi graph restricted to the edges of the observed $g$.  Exactly how the random selection of a subgraph/subtable is performed is not critical, and may be tuned to enhance performance of the random walk. Our implementation of Algorithm~\ref{alg:MH} allows the user to specify bias toward picking only directed edges or only reciprocated edges, for example.

\smallskip
{\footnotesize  
\begin{algorithm}[H]
\label{alg:MH}
\LinesNumbered
\DontPrintSemicolon
\SetAlgoLined
\SetKwInOut{Input}{Input}
\SetKwInOut{Output}{Output}
\SetKwInOut{Assumptions}{Assumptions}
\Input{$u$, a dyad classification  
table representing an observed (multi)graph $g\in\G_n$, as defined on p.\pageref{defn:loglinearERGM};\\ 
 [optional] $M$, an integer or a table of cell bounds, of same dimensions as $u$;\\ 
$A$, an integer matrix such that $Au=t(u)$,  \\ \quad where $t(u)$ is the sufficient statistics vector for the model;\\
 $N$, the number of steps of the Markov chain;\\
 $f\left(\cdot|t(u)\right)$ conditional probability distribution;\\
 $GF(\cdot)$, the test statistic.}
\Output{Estimate of  the $p$-value for the goodness of fit test\\ \quad  of  the log-linear ERGM \eqref{eq:ERGM} with sufficient statistics $t$ to the observed data $g$.} 
\BlankLine
Compute the MLE $\hat{u}$ for $u$.\;
Set $GF_{\text{observed}}:=GF(u)$.  \;  
Set $k:=0$. \;
Set  constant.fiber.flag $:=$ true. \label{step:constantflag} \;

\For{$i=1$ \KwTo $N$}{
 Randomly select a subtable $v$ of the observed table $u$ such that $v$ represents the dyad-classification table of a subgraph of $g$. \label{step:selectSubtable} \;
Construct a  Markov move $m=m(v,A)$ using Algorithm~\ref{alg:NextMove}.\; 
If $u+m\in\mathcal G_n$ is not a valid (multi) graph in the sample space, then repeat Step~\ref{step:selectSubtable}.\;
$q=\min\left\{1, \frac{f(U=u+m|t)}{f(U=u|t)}\right\}$. \label{step:acceptance}\;
$u = \begin{cases} 
u+m, & \text{ with probability } q\\
u, & \text{ with probability } 1-q\end{cases}$. \label{step:nexttable} \;
\uIf {constant.fiber.flag $=$ true and $GF\left(u\right)\neq GF_{\text{observed}}$}{ constant.fiber.flag $:=$ false.}\label{step:constantfalse}
\uIf {$GF\left(u\right)>GF_{\text{observed}}$}{   
	$k:=k+1$.}
 }
\uIf {constant.fiber.flag $=$ true}{Print: Statistic GF was constant on MCMC sample; run chain longer, or use another GF statistic.}\label{step:constantWarning} 
Output $\frac{k}{N}$.\\
\caption{Dynamic Markov bases Metropolis-Hastings for log-linear ERGMs}
\end{algorithm}
}
\smallskip
Note that if the estimated $p$-value output by the algorithm is $k/N=0$, this means $k=0$, and so no graphs were discovered with a GoF statistic value larger than the observed. There are two cases in which this can happen: either the GoF statistic itself is constant on the fiber (this includes the case when the fiber contains only one graph), or at least some of the graphs discovered have a smaller GoF value than the observed one. To distinguish the two cases, we keep track of whether the GoF statistic is constant during the run of the algorithm (Steps \ref{step:constantflag} and \ref{step:constantfalse}) and, if it is, we print a warning message in Step~\ref{step:constantWarning}.

\smallskip
{\footnotesize  
\begin{algorithm}[H]
\label{alg:NextMove}
\LinesNumbered
\DontPrintSemicolon
\SetAlgoLined
\SetKwInOut{Input}{Input}
\SetKwInOut{Output}{Output}
\SetKwInOut{Assumptions}{Assumptions}
\Input{
$v$, a subtable of the dyad classification  table representing an observed (multi)graph $g\in\G_n$; \\
$A$, an integer matrix such that $Au=t(u)$,  \\ \quad where $t(u)$ is the sufficient statistics vector for the model.\\ 
}
\Output{$m$, a table of the same format as $v$, with integer entries, such that $Am=0$.}
\caption{Construction of one Markov move}
Convert $v$ to a multihypergraph $\cR$; \label{step:convert} 
Find a multiset of hyperedges $\cB$ from $\cH_{\M}$ that balances $\cR$, ensuring that each Graver move $(\cR, \cB)$, which respects cell bounds given by $M$, if any, has  positive probability of being constructed. \label{step:getMove} \; 
 Set $m=\cR-\cB$ and convert it back to contingency table format. \; 
\end{algorithm}
}

Step~\ref{step:convert} in Algorithm~\ref{alg:NextMove} amounts to a straightforward task of translation from a log-linear model's design matrix to the corresponding hypergaph.  
Technical details of the hypergraph construction for each of the models we consider here can be found in Appendix~\ref{sec:appendix:hypergraphs}, with details of this conversion in Appendix~\ref{sec:appendix:table_to_hypergraph}, based on \cite{PShypergraph}. 
 This is easily implemented for any model, and incurs no additional complexity. 

Step~\ref{step:getMove} in Algorithm~\ref{alg:NextMove} is model specific and is the core of the dynamic Markov basis sampler. In  Section~\ref{sec:GoFknownmodels},  we discuss the implementation of this step for the specific models from Section~\ref{sec:knownmodels}. 
 When this step is done combinatorially,  implemented to respect the  assumption in the following 
 Theorem, Algorithm~\ref{alg:MH} provides a general goodness-of-fit test for the log-linear model.  

\begin{theorem}\label{thm:GoF}
Let $g_0$ be an observed graph from a log-linear ERGM with sufficient statistics vector $t(G)$  additive over the dyads of the graph, as in Definition~\ref{defn:loglinearERGM}. 
Let $\F_t(g_0) = \{ g \in \mathcal G_n \ : \ t(g) = t(g_0) \}$  be the   fiber of $g_0$ under the model, as defined in~\eqref{eq:fiber}. 
  Let $GF_0 = GF(g_0)$ be observed value of the goodness of fit statistic. 
    
Suppose that Algorithm~\ref{alg:NextMove} constructs any applicable Markov move with positive probability, and  suppose in addition that the probability of moving from graph $g_1$ to graph $g_2$ via this algorithm is equal to the probability of moving from $g_2$ to $g_1$.

Then Algorithm~\ref{alg:MH} is a Metropolis-Hastings algorithm and as $N \to \infty$ the output will, with probability $1$, 
equal the exact conditional $p$-value, defined as $P\left( GF\left(G\right)>GF_0\mid t(G)=t(g_0)\right)$ for a random graph $G$ drawn from the log-linear ERGM. 
\end{theorem} 

\begin{proof}
First, note that the random walk in Algorithm~\ref{alg:MH}  is a walk on the correct reference set for the goodness-of-fit test. Since the output of Algorithm~\ref{alg:NextMove} is a table $m$ such that $Am=0$, the new table $u+m$ constructed in Step~\ref{step:nexttable} of Algorithm~\ref{alg:MH} will be in $\F_t(g)$,  the fiber  of the observed table under the model. This follows by  the definition of the model fibers in Equation~\eqref{eq:fiber} on page~\pageref{eq:fiber}, since $Au = A(u+m)$, thus $t(u)=t(u+m)$.

By  \citet[Lemma 2.1]{DS98}, a result which applies to any log-linear model, if Algorithm~\ref{alg:NextMove} constructs any applicable Markov move with positive probability,  then the fiber $\F_t(g)$ is connected, and thus the resulting Markov chain random walk is irreducible. Note that this result, along with its algebraic proof, assumes that $M$ is \emph{not} provided. 

In case the algorithm is given the optional input $M$ on cell bounds of the dyad classification table or, equivalently, multiplicity bounds on each observable edge of the (multi)graph from the sample space $\mathcal G_n$, Markov bases from the Fundamental Theorem of \citet{DS98} do not guarantee connectedness of all fibers.  
In contrast, if $M$ is given, then every move in the Graver basis which respects cell bounds given by $M$ has a nonzero probability of being constructed by  Algorithm~\ref{alg:NextMove}. 
Connectedness of observable fibers follows from \cite[Proposition 1]{GPS16}. The proof is algebraic; we outline it for completeness:  By the fundamental theorem of Markov bases, any Markov move corresponds to a binomial in the corresponding toric ideal; this ideal is uniquely defined for each log-linear model. Combinatorial commutative algebra literature, for example \cite{St}, provides that every such binomial  can be written as what is called a conformal sum of Graver basis elements. Intuitively, a conformal sum is defined as a sum of binomials where no cancellations of terms occur; in graphs this would translate to a sequence of moves that doesn't add an edge by one move just to   remove it by another.  Sampling constraints on $\mathcal G_n$ translate to degree bounds of these binomials, and by conformality, even when classical Markov bases do not connect the constrained fibers, degree-bounded Graver bases do.  
The discussion preceding the Proposition in \citeauthor{GPS16} contains more explicit details on the algebra of this proof. 

  Graver bases elements are difficult to compute in general, but  Algorithm~\ref{alg:NextMove}  is designed to construct each such element with positive probability.    
  Therefore each $\F_t(g)$ will be connected even in presence of cell bounds. 

Aperiodicity of the chain follows from the fact that any applicable moves has   a positive probability of being constructed by Algorithm~\ref{alg:NextMove}. In fact, the chain could be aperiodic even if the algorithm constructed only a minimal set of connecting moves for a given fiber, but we construct more:  every entry of the transition matrix of this Markov chain is positive. 

Symmetry of the chain is assumed to hold in the design of the procedure in Algorithm~\ref{alg:NextMove}. 


The  claim of convergence to the conditional $p$-value now follows by  standard considerations on the  Metropolis-Hastings algorithm, for example from \cite{DSS09} or \cite{RC99}.  The acceptance ratio in Step~\ref{step:acceptance} ensures the correct stationary distribution: the conditional distribution on the fiber of the observed graph. 
\end{proof}


\citet{GPS16} use a special instance of this general framework for testing model fit of the $p_1$ model with dyad-specific reciprocation.  Their implementation of Algorithm~\ref{alg:NextMove} constructs a move at each step using a random combinatorial process.  Each element of the Graver basis has a non-zero probability of being constructed, and thus the resulting chain   is indeed irreducible. Theorem 3.8. therein proves symmetry. 
\begin{remark}[On symmetry] \label{rmk:symmetry}
	In the original Markov bases algorithm \citep{DS98}, the proposal distribution was symmetric `for free'; namely, the assumed knowledge of a full Markov basis allowed the chain to pick one move at random, and then  either apply it or its negative. Thus the chain was reversible by design. However, the main motivation of constructing Markov moves dynamically is the computational bottleneck of obtaining the full Markov basis in the first place. This is the reason we added the symmetry assumption in the statement of \autoref{thm:GoF}. 	When this holds,  the Markov chain in Algorithm~\ref{alg:MH} is also reversible, and the additional multiplier by the proposal distribution does not appear in the Hastings acceptance ratio. This setup resembles Algorithm 1.1.13 in \cite{DSS09}, which is the full statement of the algorithm as proposed by \citeauthor{DS98}, also not including the additional multiplier in the acceptance ratio since the move proposal was symmetric. 
\end{remark} 

In light of Remark~\ref{rmk:symmetry}, the fact that we build our implementation from the algorithm primitive implemented in \cite{GPS16} implies that our construction inherits symmetry. This is stated formally in Proposition~\ref{prop:symmetryForOurModels}.

A note on mixing  of the Markov chain is in order. Ideally, Step~\ref{step:getMove} in Algorithm~\ref{alg:NextMove}   should take advantage of the specific structure of the hypergraph. If only applicable moves are constructed by design---and this is a nontrivial task to solve in general, but feasible for specific instances of the problem---then the rejection  step from the usual Metropolis-Hastings proposed by  \citeauthor{DS98} is bypassed. 
 This, in turn, should have positive impact to the mixing time of the chain. An important example of this was demonstrated by \citeauthor{GPS16}, whose combinatorial sampler vastly outperformed \citeauthor{OHT}'s sampler on a $\beta$-model fiber a small graph. We  discuss related work on mixing time in Section~\ref{sec:conclusion}.

\subsection{Examples of log-linear ERGMs: instances of the general algorithm} \label{sec:GoFknownmodels}

 While Algorithm~\ref{alg:MH} applies to any log-linear ERGM, we demonstrate the general methodology on the $\beta$-model (see \cite{CDS11}
 ), the $p_1$-model with three different reciprocation effects~\cite{HL81,FW81}, and block versions of these. 
 Crucially, we also implement the GoF test  on versions of these models that pose structural zeros, discussed by \citet[Section 5.1]{bishop2007discrete}, a feature that greatly complicates construction and application of Markov moves in practice.  Already \citeauthor{DS98}, and the subsequent literature including \cite{SteveAleMe-holland}, have acknowledged the difficulty of applying Markov bases when there are sampling constraints or structural zeros  on a fiber; naturally, such a structure leads to many forbidden moves and fiber connectivity using the classical sampler built on Markov bases is no longer guaranteed. 

 There are two model-dependent steps that need to be specified for each specific instance of the general algorithm: the choice of a GoF statistic in Algorithm~\ref{alg:MH}, and the implementation of dynamic move generation in Step~\ref{step:getMove} in Algorithm~\ref{alg:NextMove}.

\paragraph{Choice of a GoF statistic.} In this section, we choose the $\chi^2$ statistic, \(
	\sum_{1\leq i,j\leq n} \frac{(\hat g_{ij}-g_{ij})^2}{\hat g_{ij}}.
\)
This is a reasonable choice for model fitting log-linear models, as outlined in Fienberg's early work; see \cite{PRF:09} for a summary. 
The choice of a GoF statistic cannot be universal, however, because it can happen that it is constant on non-trivial fibers, in which case the test for that model, or at least that fiber, is vacuous. The only known examples so far where the $\chi^2$ statistic degenerates to a constant are  the classical 
  blockmodel and, consequently, the  Erd\"os-Renyi model itself. These examples are discussed by \citet[Section 3.3]{karwa2016exact}, along with proposals for alternative  $\chi^2$-like GoF statistics. 

It should be noted since we are using the  $\chi^2$ statistic, we compute the MLE $\hat g_{ij}$ at the start of the test and then at each step of the walk we compare the current graph against the MLE. Utilizing the connection to log-linear models, we compute the MLE by standard IPS algorithms for contingency tables, for example, using {\tt loglin} in {\tt R},  
 and thus avoid the use of MCMC-MLE algorithms for general ERGMs (cf. \cite[Section 5.1]{Hunter}). 

\paragraph{Dynamic Markov moves.} 
Herein we offer the general strategy of how we implement dynamic move generation, that is, Step~\ref{step:getMove} in Algorithm~\ref{alg:NextMove}. 
 For a complete description of the parameter hypergraphs for each model see  Appendix~\ref{sec:appendix:hypergraphs}.
 The parameter hypergraph of the $\beta$-model is the simplest, since it is isomorphic to the complete graph on $n$ vertices, where $n$ is the number of nodes in the observed network. Thus, after the graph conversion step from Appnedix~\ref{sec:appendix:table_to_hypergraph},  we use a variation of the classic edge swap algorithm (\cite{kannan1999simple}, \cite{rao1996markov}, \cite{ryser1987combinatorial}, \cite{taylor1981contrained}, \cite{tabourier2011generating}, \cite{wang2020fast}).  
 Namely, we sample a set of edges $E_1=\{\{u_1,v_1\}, \ldots \{u_k, v_k\}\}$ from the current network state and then use $E_1$ to form a closed even walk by adding the edges $E_2 = \{\{u_1,v_2\}, \{u_2,v_3\}, \ldots \{u_k, v_1\}\}$. We then replace the edges in $E_1$ by the edges in $E_2$, and if the resulting network is simple, we move from the current network state to a new network state, else the move is rejected.  This chain is, by design, symmetric  and ergodic. 
\begin{figure}[!h]
\centering
\begin{tikzpicture}[scale=.6, every node/.style={minimum size=5mm, text centered}]
   \foreach \i/\color in {1/blue, 2/blue} {
       \node[draw, circle, preaction={fill=\color, fill opacity=0.1}, draw=\color, inner sep=0pt] (n\i) at ({(-\i)*360/4+180}:1.7) {{\i}};
   }
   \foreach \i/\color in {3/red, 4/red} {
       \node[draw, circle, preaction={fill=\color, fill opacity=0.1}, draw=\color, rectangle, inner sep=0pt] (n\i) at ({(-\i)*360/4+180}:1.7) {{\i}};
   }
\tikzset{edge/.style = {->,> = stealth'}}
   
	\draw[edge] (n1) to (n2);
	\draw[edge] (n2) to (n1);
	\draw[red, thick] [edge] (n1) to (n3);
	\draw[edge] (n1) to (n4);
	\draw[red, thick] [edge] (n4) to (n2);
	 \node (edgesE1) at (0,-3) {$E_1=\{\{1,3\}, \{4, 2\}\}$};
\end{tikzpicture}
\quad
\begin{tikzpicture}[scale=.6, every node/.style={minimum size=5mm, text centered}]
\tikzset{edge/.style = {-,> = stealth'}}
	 \node (n1) at (-2,1.5) {1};
	\node (n2) at (2,-1.5) {2};
	 \node (n3) at (-1.5,-1.5) {3};
	 \node (n4) at (1.5,1.5) {4};
  
	\draw[red, thick] [edge] (n1) to (n3);
	\draw[red, thick] [edge] (n4) to (n2);
	\draw[blue, thick] [edge] (n1) to (n2);
	\draw[blue, thick] [edge] (n4) to (n3);
	 \node (walk) at (0,-3) {The walk primitive};
\end{tikzpicture}
\quad
\begin{tikzpicture}[scale=.6, every node/.style={minimum size=5mm, text centered}]
   \foreach \i/\color in {1/blue, 2/blue} {
       \node[draw, circle, preaction={fill=\color, fill opacity=0.1}, draw=\color, inner sep=0pt] (n\i) at ({(-\i)*360/4+180}:1.7) {{\i}};
   }
   \foreach \i/\color in {3/red, 4/red} {
       \node[draw, circle, preaction={fill=\color, fill opacity=0.1}, draw=\color, rectangle, inner sep=0pt] (n\i) at ({(-\i)*360/4+180}:1.7) {{\i}};
   }
\tikzset{edge/.style = {->,> = stealth'}}
   
	\draw[edge] (n1) to (n2);
	\draw[edge] (n2) to (n1);
	\draw[blue, thick, bend left=40] [edge] (n1) to (n2);
	\draw[edge] (n1) to (n4);
	\draw[blue, thick] [edge] (n4) to (n3);
	 \node (edgesE2) at (0,-3) {$E_2 = \{\{1,2\}, \{4,3\}\}$};
\end{tikzpicture}
\vspace{-3mm}
\caption{One move proposed by the edge-swap algorithm primitive.}  
\label{fig:a random move}
\vspace{-4mm}
\end{figure}

 We  refer to our variations of the classic edge swap algorithm as the \emph{edge-swap algorithm primitive}. \label{edge-swap-primitivie}
 It is a basic building block for all of the samplers we have implemented. 
An example of one run of the edge-swap algorithm primitive for the graph from Figure~\ref{fig:example-graph} is shown in Figure~\ref{fig:a random move}. It is important to note that this algorithm primitive is a building block for the moves necessary to sample from various models. To understand the complete algorithm instance for each model, one needs to understand the model parameter hypergraphs.

The parameter hypergraph $\mathcal H$ of the dyad-specific $p_1$-model has both the complete graph on $n$ vertices $K_n$ and the complete bipartite graph $K_{n,n}$ with the $(i,i)$th edge removed as induced subgraphs. These induced subgraphs correspond to the directed and reciprocated edges, respectively. To preserve both the degree of  each node and the total number of reciprocated edges incident to each node, the network is first split into directed and reciprocated edges, and then we use the variation of the classic edge swap algorithm described above to move among subgraphs with a fixed degree sequence of $K_n$ and $K_{n,n}\setminus\{ (i,i) : 1 \leq  i\leq n\}$. Combining two moves, one for each induced subgraph, gives us a move on the space of all multiset of edges of $\mathcal H$ with the prescribed fixed degree sequence. This is the symmetric and ergodic chain described in \cite{GPS16}. 

For the zero-reciprocation  variant of the $p_1$-model, we see that $K_{n,n}\setminus\{ (i,i) : 1 \leq  i\leq n\}$ is again an induced subgraph of the parameter hypergraph $\mathcal H$. Each reciprocated edge $i\leftrightarrow j$ corresponds to a hyperedge of size four in $\mathcal H$, thus we first divide each of these hyperedges into two edges of size two, one which corresponds to the configuration $i\leftarrow j$ and the other to $i\rightarrow j$; this is akin to uncoupling the reciprocated edge $i\leftrightarrow j$ into two directed edges, $i\leftarrow j$ and $i\rightarrow j$. After this uncoupling, we then generate a degree-sequence preserving move on $K_{n,n}\setminus\{ (i,i) : 1 \leq n\}$. The constant-reciprocation variant is implemented similarly to the zero reciprocation one with an additional step to ensure that the number of reciprocated edges in the final move is preserved. This step makes sure the vertex corresponding to $\rho$ (see Appendix \ref{sec:hypP1zero}) 
 is covered the same number of times by both multisets of hyperedges.
Each of these samplers proposes a combination of directed and/or reciprocated moves, as required, and then ensures that the proposed move is applicable by checking for edge conflicts with the current network. 

The samplers for $\beta$-SBM and $p_1$-SBM models also utilize the extension of the classic edge-swap algorithm primitive described above with the additional constraint of preserving the sufficient statistics dictated by the blocks.   
Finally, structural zeros for any of the models are handled by  checking whether the proposed new network attempts to place an edge over a structural zero; in that case, a new proposed move is constructed. 

\smallskip 
To increase the efficiency of the sampling algorithm, we also experimented with optional inputs that modify the chains above, but in a way that does not change the limiting distribution.  For example, when the graph is dense, we found that we are able to explore the space faster by adding a tuning parameter to the algorithm that controls the frequency of using `small' moves,  
 set to be a value between 0 and 1 that controls for the percentage of moves of size four or less.  The chain is guaranteed to remain connected as long as this parameter 
  $<1$, however, one needs to be careful, as the closer the value is set to one, the longer the chain may take to mix.  In a similar vein, for the dyad-specific $p_1$ model, there is a tuning parameter 
   vector of length 3 that controls the probabilities of selecting a move only on the directed component, selecting a move only on the reciprocated component, and selecting a combined move.  This parameter can be helpful when one component is dense and the other is sparse, as moves are easier to construct on sparser graphs, having fewer rejections. For graphs with non-empty directed and reciprocated components, the chain remains connected as long as each entry of the vector 
    is greater than zero.  Furthermore, there is a tuning parameter 
    that can be used to increase efficiency when testing fit for the $p_1$ model with constant reciprocation by allowing the user to use the  algorithm for $p_1$ with dyad-specific reciprocation, which is more efficient, for a given proportion of the moves. Finally, there  is a tuning parameter for the $\beta$-SBM model that can favor small degree-preserving moves of particular form; namely, moves that allow degrees of individual vertices to change within a block and between two blocks, and do not change the number of edges within each block or between the two blocks. 
    


\smallskip 
The next result is a special case of \autoref{thm:GoF}: it spells out that we can prove symmetry of the combinatorial implementation 
 of Step~\ref{step:getMove} in Algorithm~\ref{alg:NextMove} 
 for the models from Section~\ref{sec:models}. 
\begin{proposition}\label{prop:symmetryForOurModels}
	The output of Algorithm~\ref{alg:MH} will, with probability $1$ as $N \to \infty$, 
equal the exact conditional $p$-value for the test of goodness of fit for the $p_1$-SBM and its specializations, including all variants of the $p_1$ model and the $\beta$-SBM. 
\end{proposition}
\begin{proof}
The one item that is further specified from the main result   \autoref{thm:GoF} is the move samplers, that is,  Step~\ref{step:getMove} in Algorithm~\ref{alg:NextMove}.
The  implementation of this step for each model defined in Section~\ref{sec:models} 
is described in the text above. 
By design, each of the Markov move samplers  are all built on  the edge-swap algorithm primitive, defined on page~\pageref{edge-swap-primitivie}. 
This base algorithm, along with the structural decomposition of the model hypergraphs described in Appendix~\ref{sec:appendix:hypergraphs}, implies that our construction inherits symmetry in the same way that \citeauthor{GPS16}'s sampler did. 
The key argument in the proof of symmetry, contained 
in proof of Theorem 3.8 in that work, requires that 
a proposed  move from $g_1$ to $g_2$ adds and removes the exact same number of edges. 
 This  constraint holds for each of the samplers  described above. 
 \end{proof}


\section{Performance results}
\label{sec:applications}

We summarize the goodness-of-fit test's performance on  simulated data in Section~\ref{sec:simulateddata}.
Our method is illustrated on two experimental datasets in Sections~\ref{sec:neuronal} and \ref{sec:protein}: a neuronal network \citep{Varshneyetal11} and a protein-protein interaction network \citep{AIMC2011}. Both networks have interested applied researchers for over a decade. The experimental results  in Section~\ref{sec:applications} 
show that none of the variants of the $p_1$ fit the directed part of the neuronal data, meaning that the edge formation in these neuronal networks is not mainly driven by the attractiveness and expansiveness of nodes (neurons). 
 But a run of edge-dependent $p_1$ for the union of directed and undirected  neuronal network gives $p$-value $>0.3$. Interestingly, computations also  show that  the $p_1$-model with structural zeros fits the protein-protein network pretty well; to determine whether this fit is an artifact of the data collection process remains to be addressed in further work. 
 
 The {\tt R} code to reproduce examples from this paper can be found at \url{https://github.com/p1org/DMBHGraphs/}.

\subsection{Simulated networks}\label{sec:simulateddata}




\begin{figure}
\begin{subfigure}[b]{.31\linewidth}
\centering
\includegraphics[width=.45\linewidth]{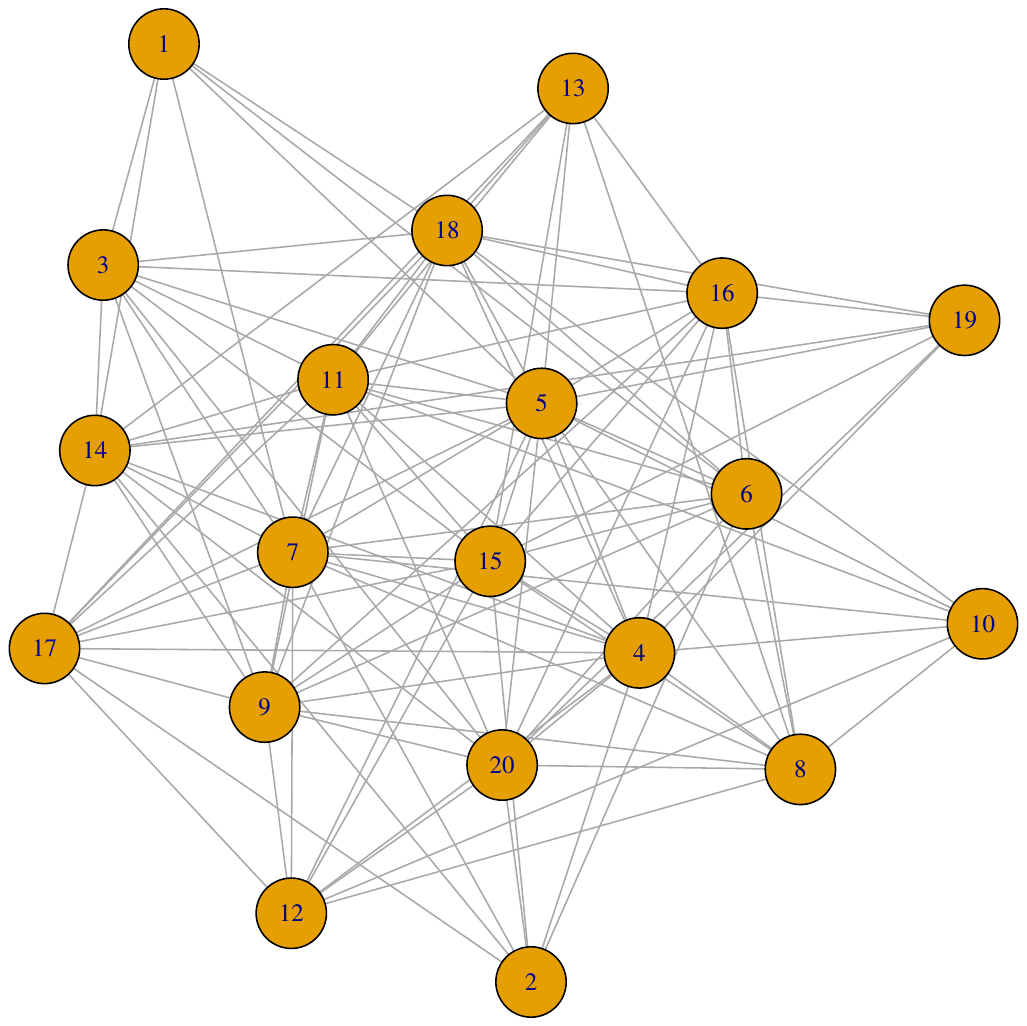} 
\includegraphics[width=.45\linewidth]{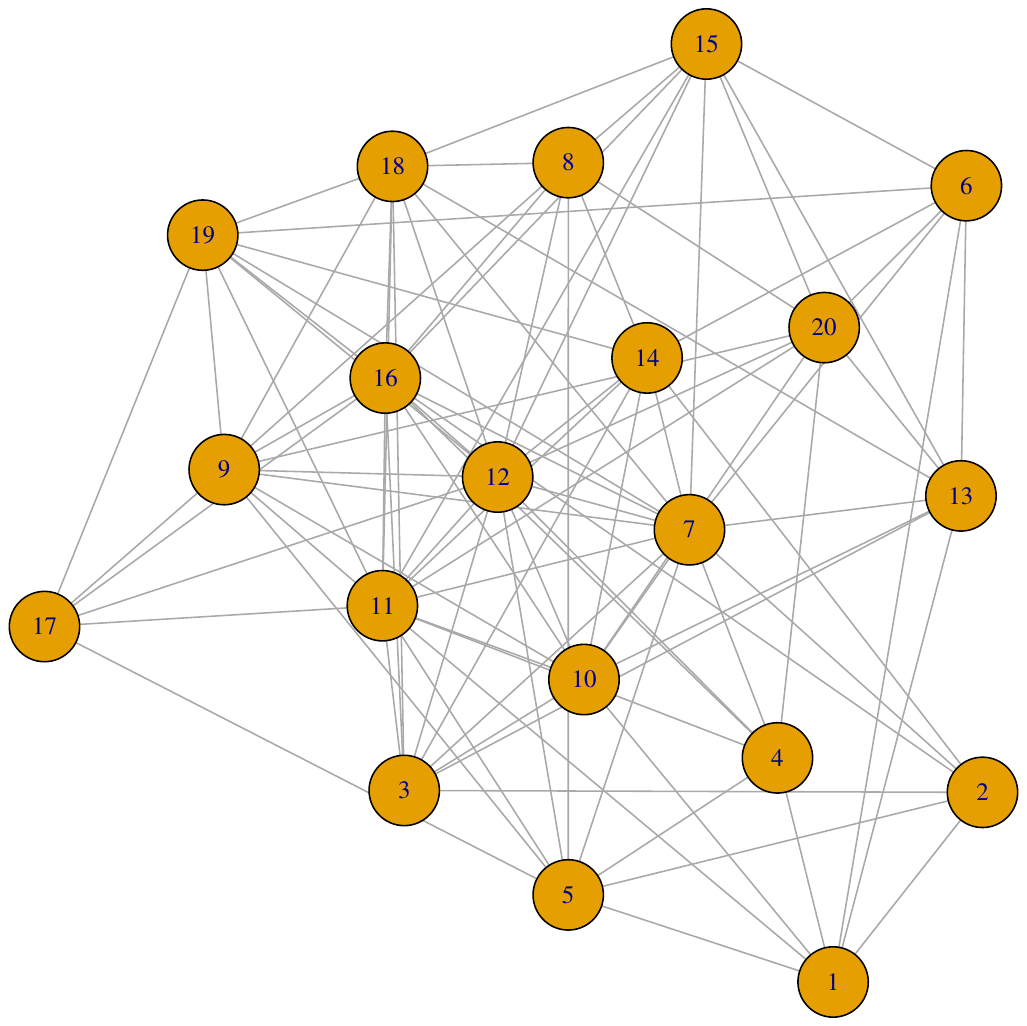} 
\caption{Two graphs simulated from the $\beta$ model on $n=20$ nodes. 
 The node degrees are heterogeneous and there is no apparent block structure in the networks.}
\label{fig:beta.simulated}
\end{subfigure}
\quad
\begin{subfigure}[b]{.31\linewidth}
\centering
\includegraphics[width=.45\linewidth]{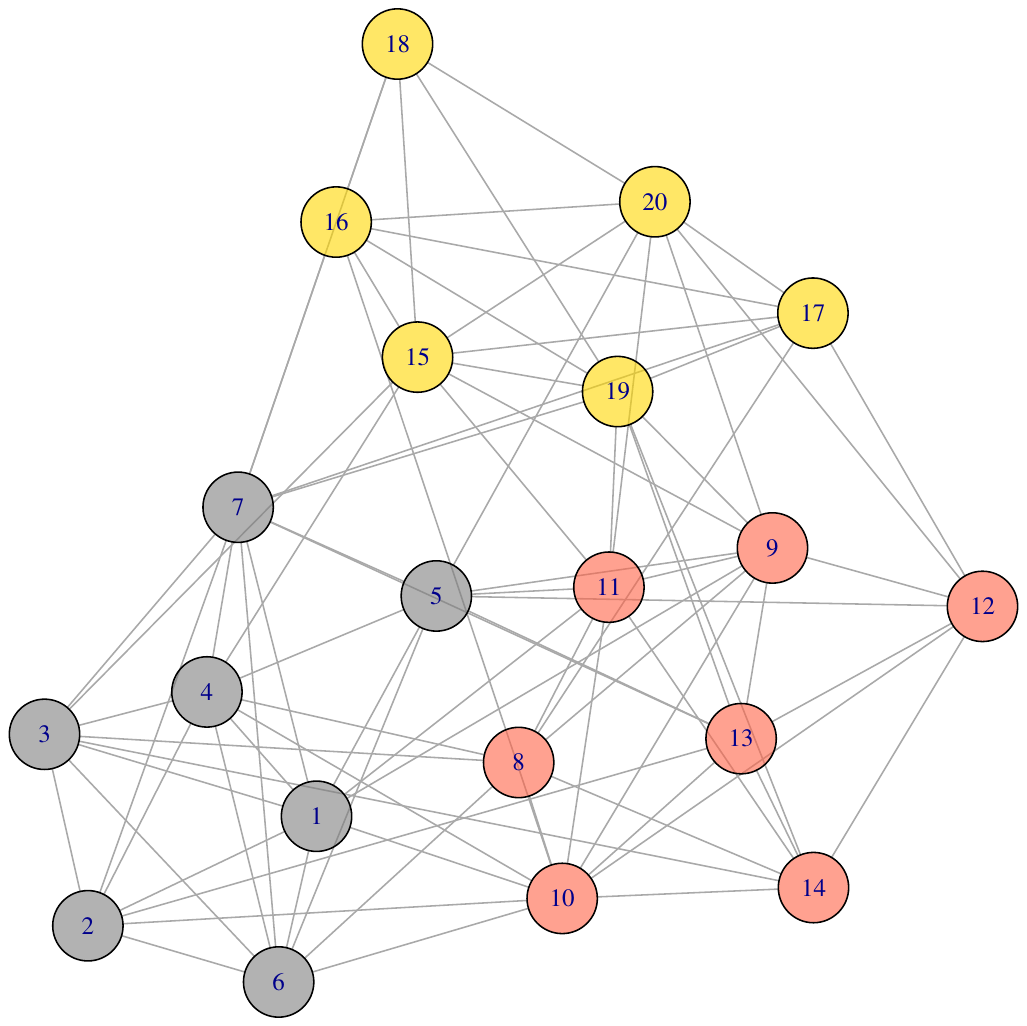} 
\includegraphics[width=.45\linewidth]{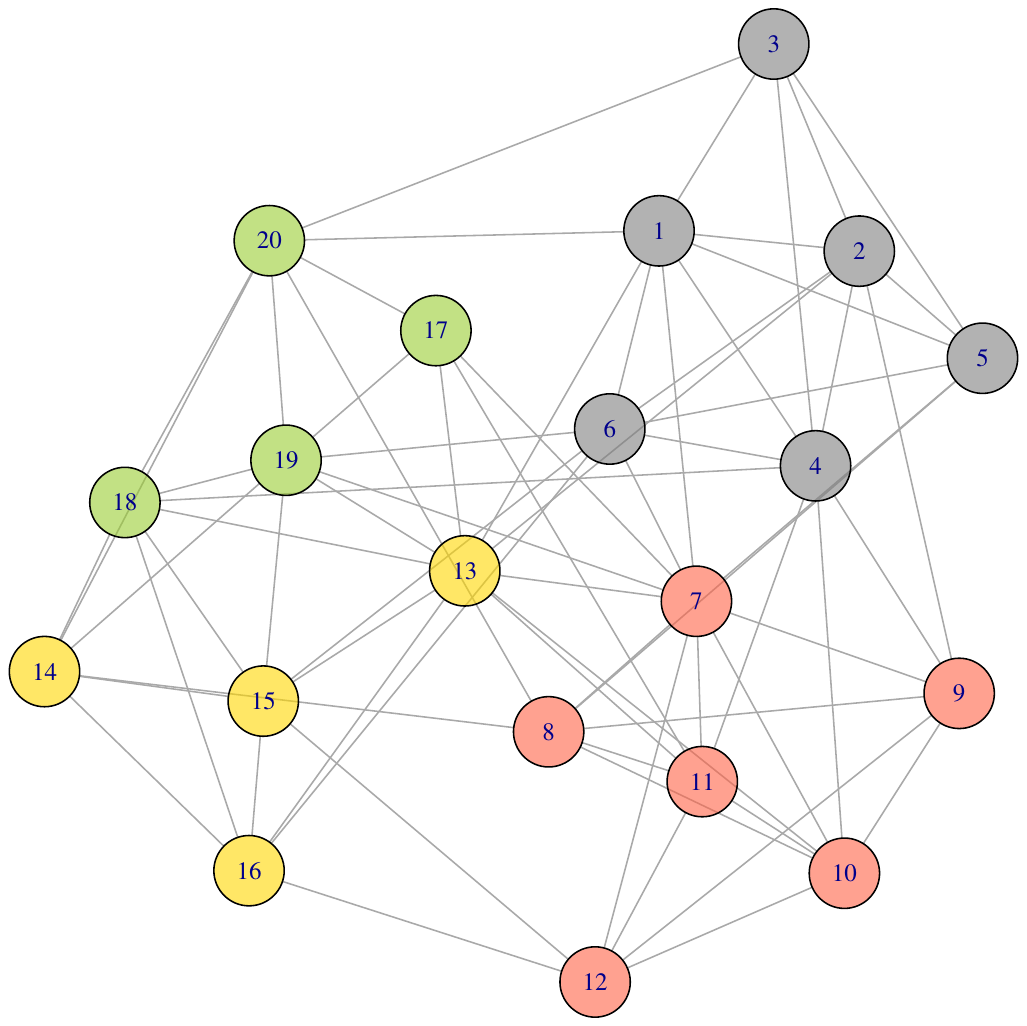} 
\caption{Two graphs simulated from the ER-SBM model on $n=20$ nodes. $k=3$ and $k=4$, respectively. Node degrees appear quite homogeneous.}
\label{fig:erSBM.simulated}
\end{subfigure}
%
\quad
\begin{subfigure}[b]{.31\linewidth}
\centering
\includegraphics[width=.45\linewidth]{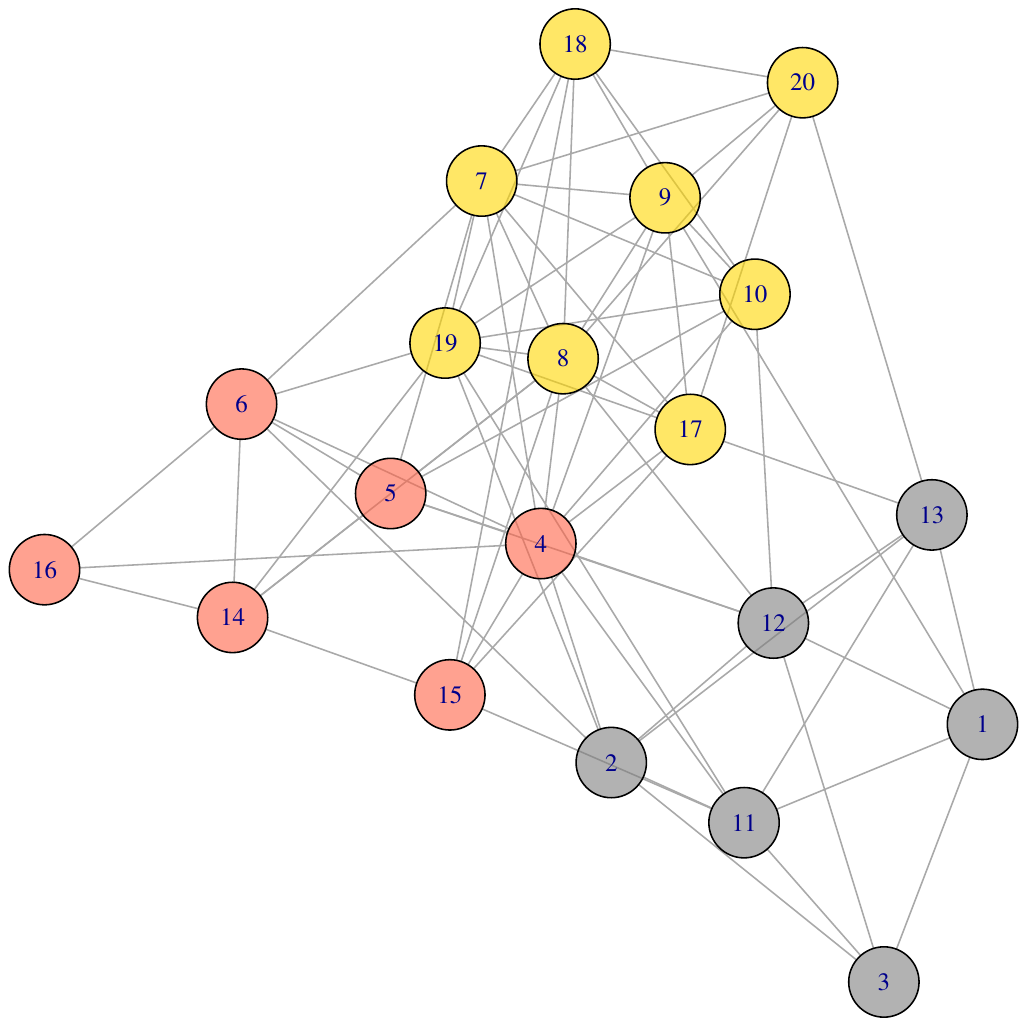} 
\includegraphics[width=.45\linewidth]{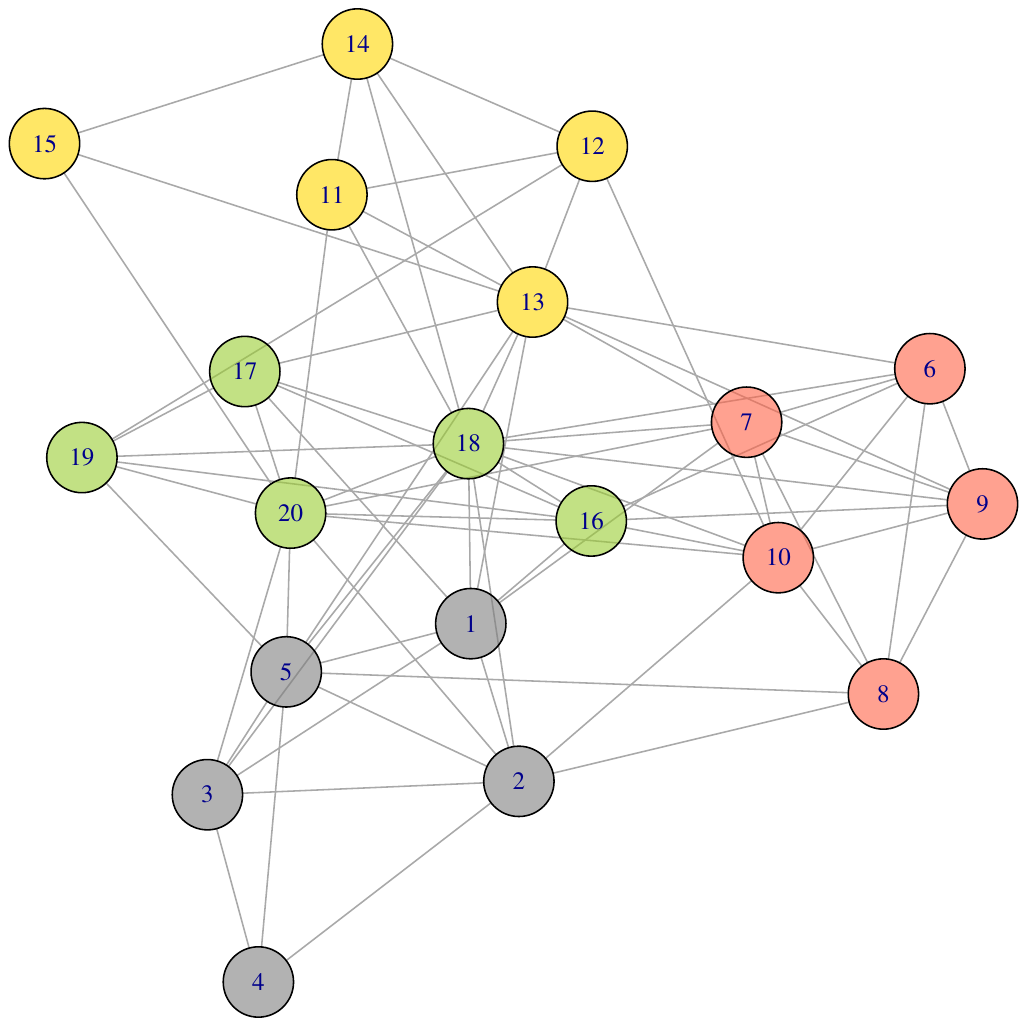} 
\caption{Two graphs simulated from the $\beta$-SBM model, $n=20$, $k=3$ and $k=4$, resp. 
Within each block, the subgraph resembles one from a $\beta$ model; otherwise node degrees appear less uniform.}
\label{fig:betaSBM.simulated}
\end{subfigure}
\caption{Graphs simulated from the various undirected models considered in this paper.}
\end{figure}

To test the power of the method, we tested the fit of the $\beta$-SBM for $100$ graphs on $n=100$ nodes simulated from a more complex model.
We chose the $\beta$-SBM since tests of goodness of fit 
  for the SBM \cite{karwa2016exact}  and some variants of the $p_1$ model \cite{GPS16} appear in the literature. 
 The simulated graphs were generated using the {\tt R} package {\tt ergm} \cite{ergm}. We simulated from the ERGM specified by the model terms {\tt degree}, {\tt nodemix}, and {\tt triangle}; in terms sufficient statistics, these model terms correspond to the sufficient statistic of the $\beta$-SBM with the number of triangles appended, and thus the $\beta$-SBM is a proper submodel of the chosen ERGM. 
The parameters are defined as follows: 
 $\beta_1,\dots,\beta_n$ are the node parameters controlling node degrees, $\alpha_{ij}$ for $1\leq i,j\leq K$ are the block connection parameters, and $\theta$ is the triangle parameter.   
Parameter values are chosen uniformly at random for the simulation study, following \cite{Lei16} (cf. \cite{karwa2016exact}). 
Of the 100 graphs, the $\beta$-SBM model was rejected, with $p$-values smaller than the nominal $0.05$,  for 90 graphs.  For each run, the Markov chain was relatively short with 10,000 steps. 
\begin{center}
\begin{tabular}{ c|c|c|c} 
number of nodes & node parameters& block parameters & triangle parameter \\
 \hline
 100 & $\beta_i\sim \Unif(-1,1)$ &  $\alpha_{ij} \sim \Unif(-1,1)$ &  $\theta\sim \Unif(-1,1)$ \\ 
 \hline
\end{tabular}
\end{center}
\begin{center}
\begin{tabular}{ c|c|c|c} 
model tested & length of Markov chain & $p$-values less than $0.05$ &  \\
 \hline
 $\beta$-SBM & 10,000 steps & 90/100 &  \\ 
 \hline
\end{tabular}
\end{center}

  Of the remaining 10 graphs where the model was not rejected, we ran longer chains to address non-mixing issues. Increasing the length of the walk to $30,000$ steps only changed the outcome for 1 of the 10 graphs. 
We note that for the $100$ sampled networks in this section, the number of triangles ranged from 39,000 to 46,000, with over 95\% of the data above 41,000.  In the 10 graphs for which the short-walk failed to reject the $\beta$-SBM model, the count was 42,000 to 44,000, with the majority at 42,000 triangles. In other words,  the triangle signal was present, but was relatively low compared to the other 90 graphs in the sample.

The one graph that  did attain a smaller $p$-value, after a longer $30,000$-step chain also showed   improvement of mixing. The $p$ value dropped  to $0.14$ from the range $0.3-0.8$. 


\subsection{Neuronal network data}\label{sec:neuronal} 

The neuronal network data set is from \cite{Varshneyetal11} and available online from the WormAtlas \url{http://www.wormatlas.org/}.  The full data set is a reconstruction of the connectome of the hermaphrodite {\it C. elegans} worm.  It contains information on 279 of the known 302 {\it C. elegans} neurons.  In the data set, edges represent chemical or electrical connections between the neurons.  The chemical connections are synaptic and directionality can be detected, so this subnetwork, which we will refer to as the \emph{chemical subnetwork}, is represented as a directed graph.  The electrical connections are recorded without direction, thus, this subnetwork, which we will refer to as the \emph{electrical subnetwork}, is represented as an undirected graph. As the authors in \cite{Varshneyetal11}, we will analyze the complete network (the union of the chemical and electrical subnetwork), as well as each subnetwork individually.

The data set also contains vertex attributes that we will use to test for block effects.  The attributes that we will focus on are functional classification, which sorts the neurons into three types of neurons, sensory neurons, interneurons, and motor neurons; regional, which classifies neurons according to whether they are found in the head, mid-body, or tail of the worm; and ganglion group (AY NeuronType), a specification that partitions the neurons into 10 groups as described by \cite{AY92}.


\paragraph{Simulation results.} 

\begin{figure}
\begin{subfigure}[b]{1\linewidth}
\centering 
\includegraphics[width=.35\linewidth]{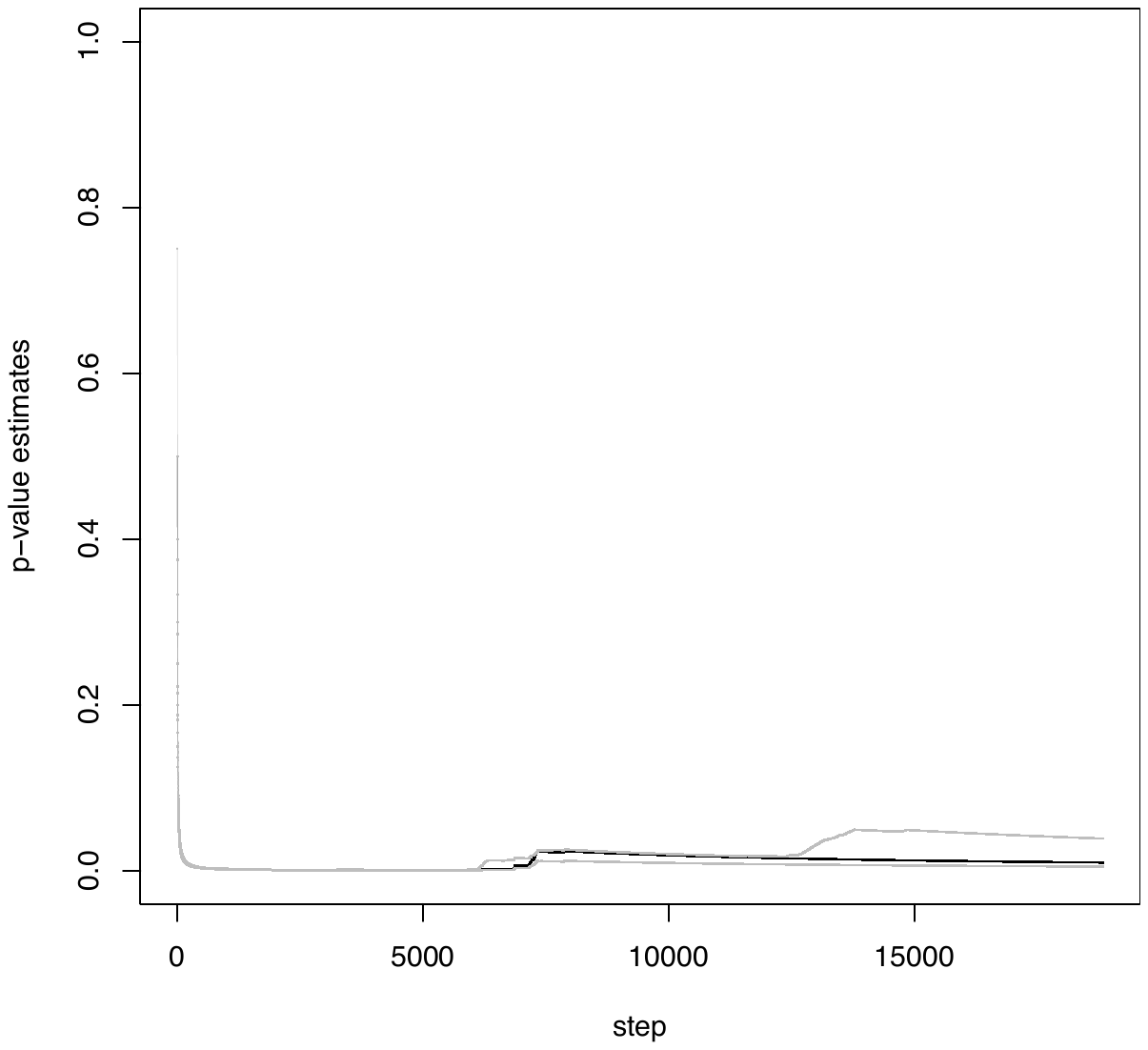}
\includegraphics[width=.35\linewidth]{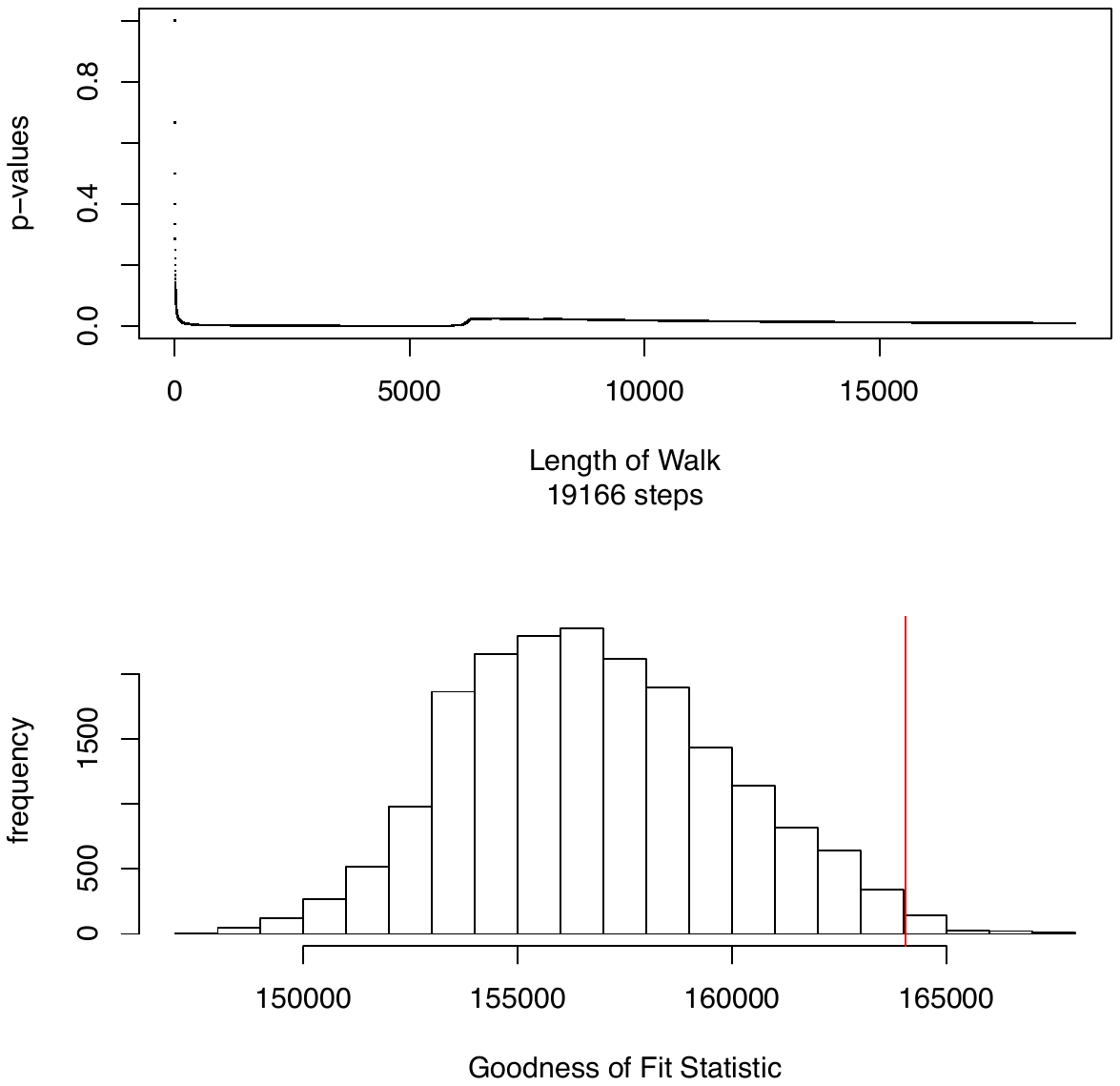}
\caption{Test of model fit for the dyad-specific $p_1$ model on the neuronal chemical (directed) network. Left: $p$-value quartiles for $3$ iterations of the Markov chain. 
  Right: typical $p$-value estimate 
   and the sampling distribution of the goodness-of-fit statistics (chi-square) along with its observed value.}
\label{fig:p1dyadNeuronalDirected}
\end{subfigure}

\begin{subfigure}[b]{1\linewidth}
\centering 
\includegraphics[width=.35\linewidth]{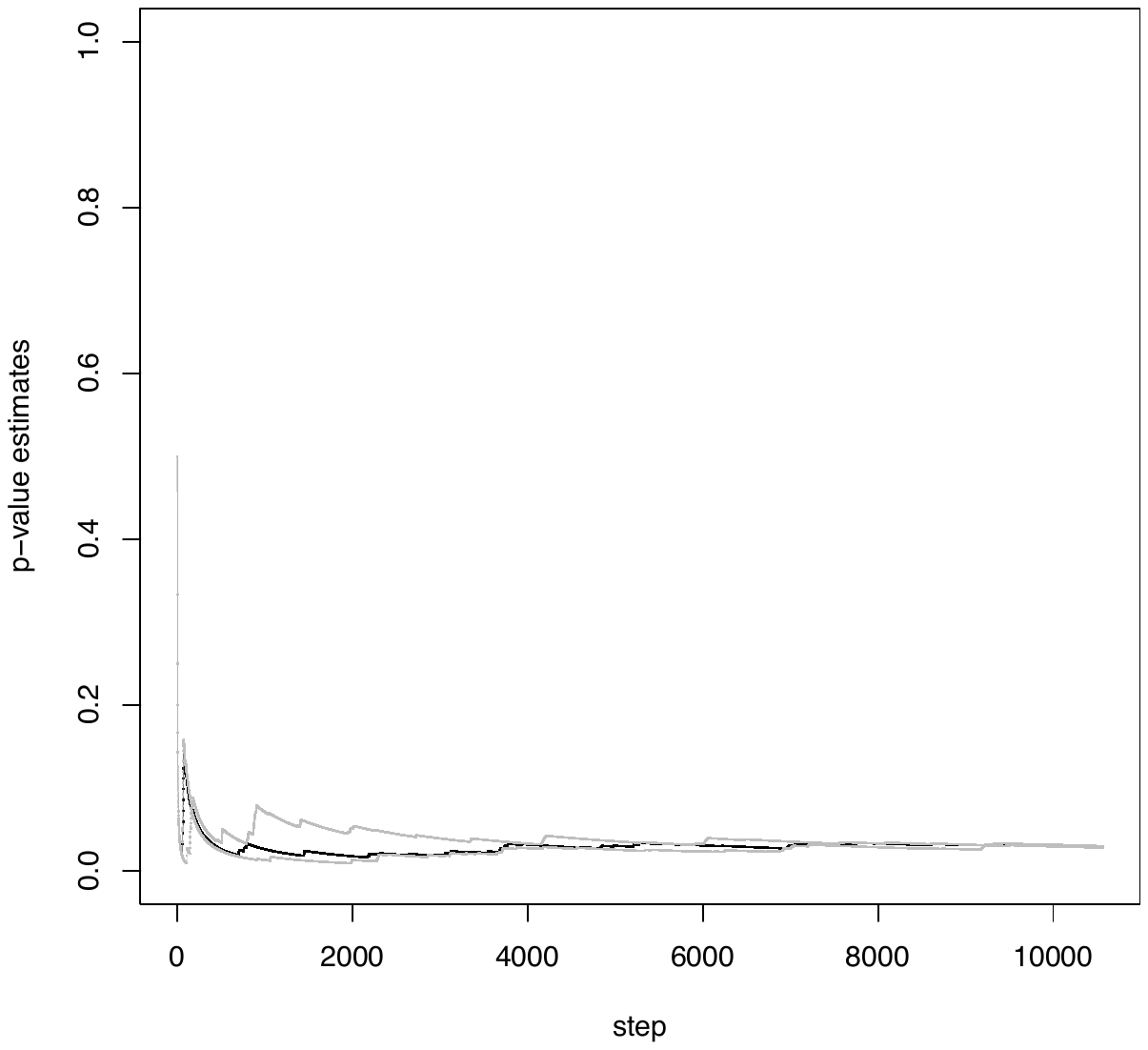}
\includegraphics[width=.35\linewidth]{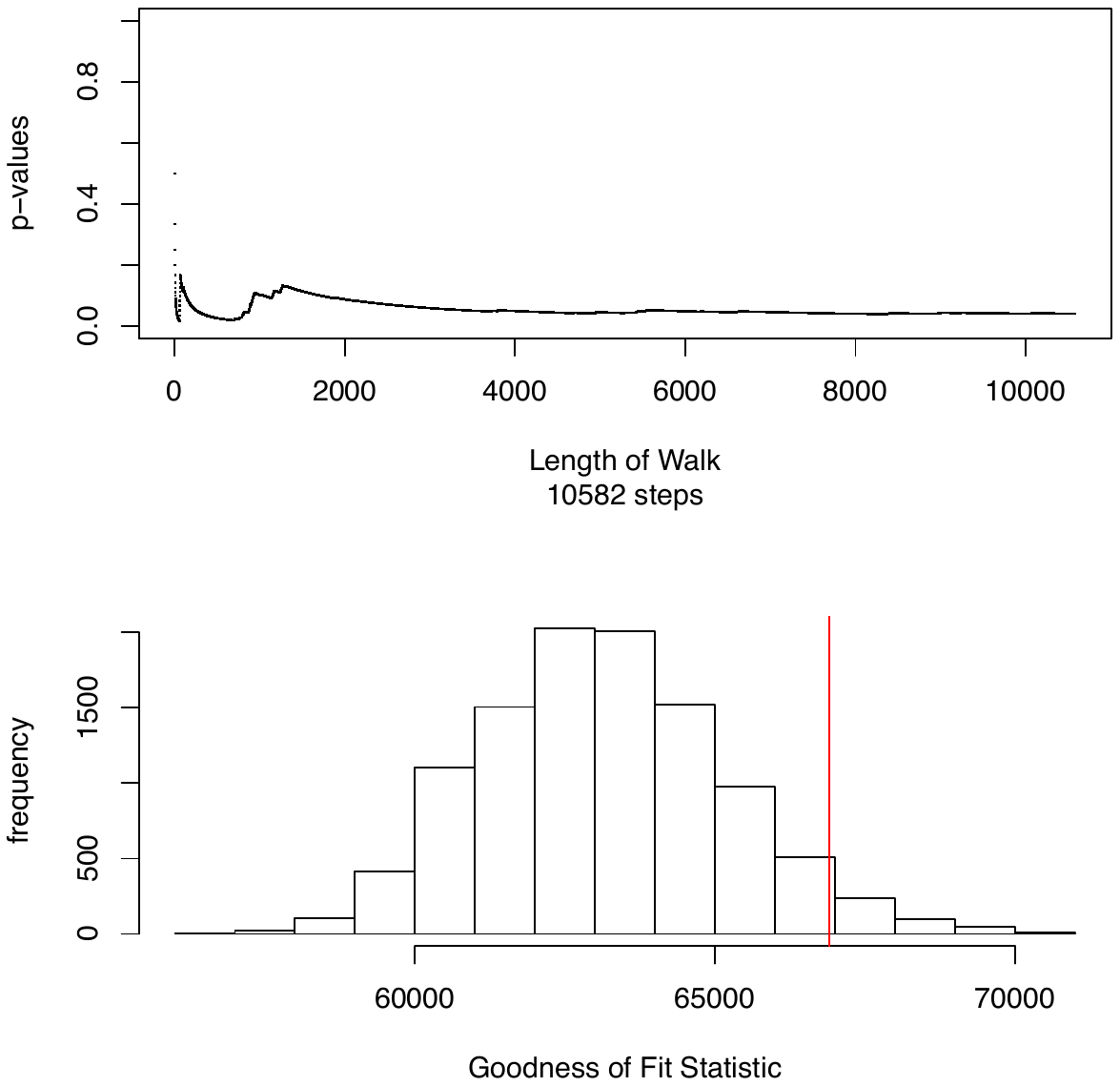}
\caption{Test of model fit for the dyad-specific $p_1$ model on the neuronal electrical (undirected) network. Left: $p$-value quartiles for $5$ iterations of the Markov chain. 
 Right: typical $p$-value estimate 
  and the sampling distribution of the goodness-of-fit statistics (chi-square) along with its observed value.}
\label{fig:betaNeuronalUndirected}
\end{subfigure}
\caption{$p_1$ model fit with dyad-specific reciprocation on two neuronal networks.}
\end{figure}

For the chemical subnetwork of the neuronal network, we tested model fit for the three variants of the $p_1$ model: zero, non-zero constant, and dyad-specific reciprocation. All three variants were rejected for simulations with as many as $100,000$ steps in the Markov chain. For illustration purposes, Figure~\ref{fig:p1dyadNeuronalDirected} shows a typical output of a shorter run for the dyad-specific reciprocation variant; note that the result is obtained in $<20,000$ steps. 
For the electrical subnetwork of the neuronal network, Figure~\ref{fig:betaNeuronalUndirected} shows the results of testing the $\beta$ model, which is  rejected at significance level $>0.04$ (with $p$-values between 0.019 and 0.04 over five iterations of the fit test).

The fact that all four models are rejected with $p$-values often significantly smaller than $0.05$ gives evidence against edge formation based on the attractiveness and expansiveness of the nodes alone, even after taking into account possible reciprocation effects.  This supports the observation in \cite{Varshneyetal11} against a scale-free model of generation and is contrary to scale invariance hypothesis suggested in \cite{RichClub}.

\begin{figure}
\begin{subfigure}[b]{1\linewidth}
\centering 
\includegraphics[width=.35\linewidth]{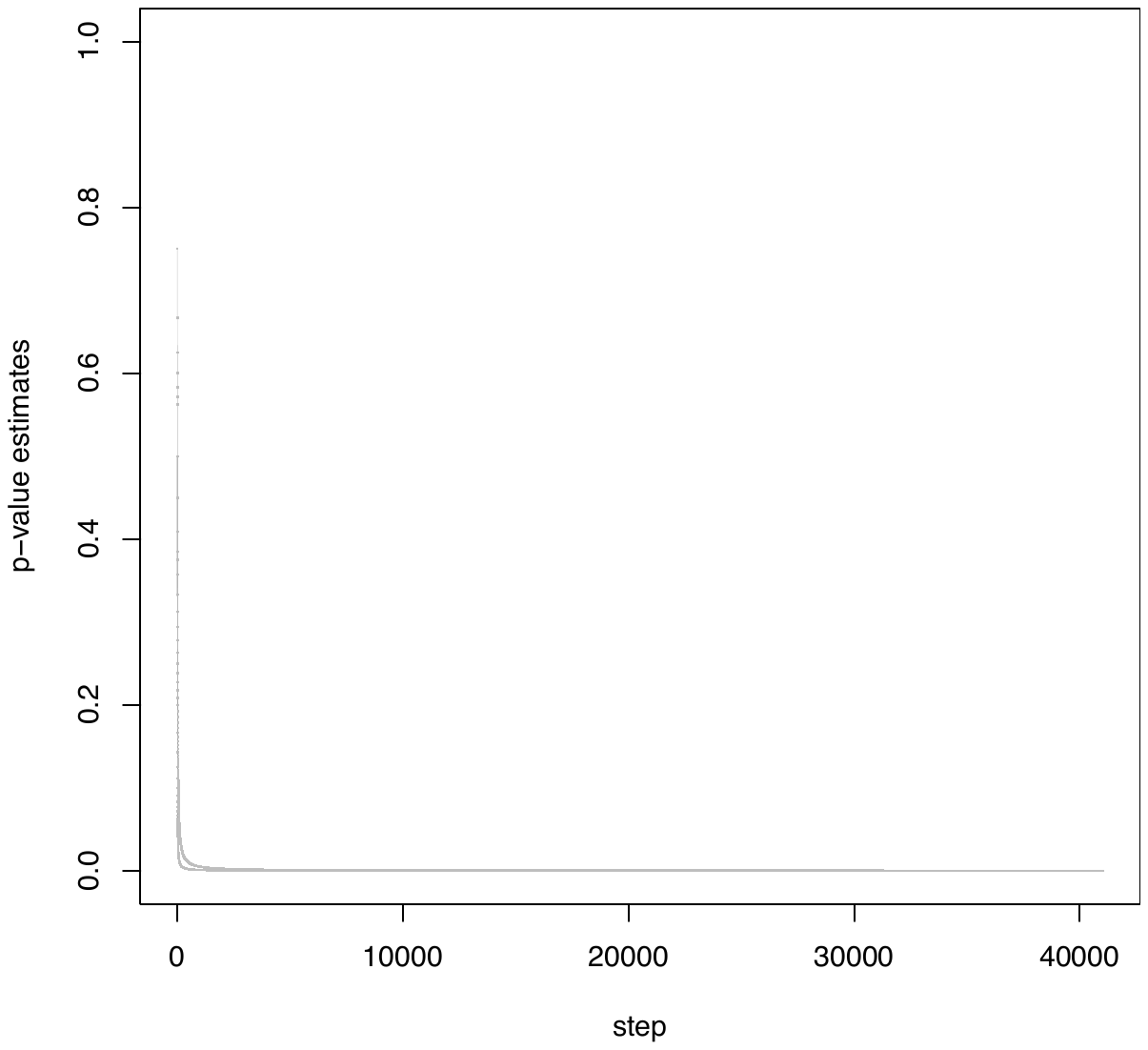}
\centering 
\includegraphics[width=.35\linewidth]{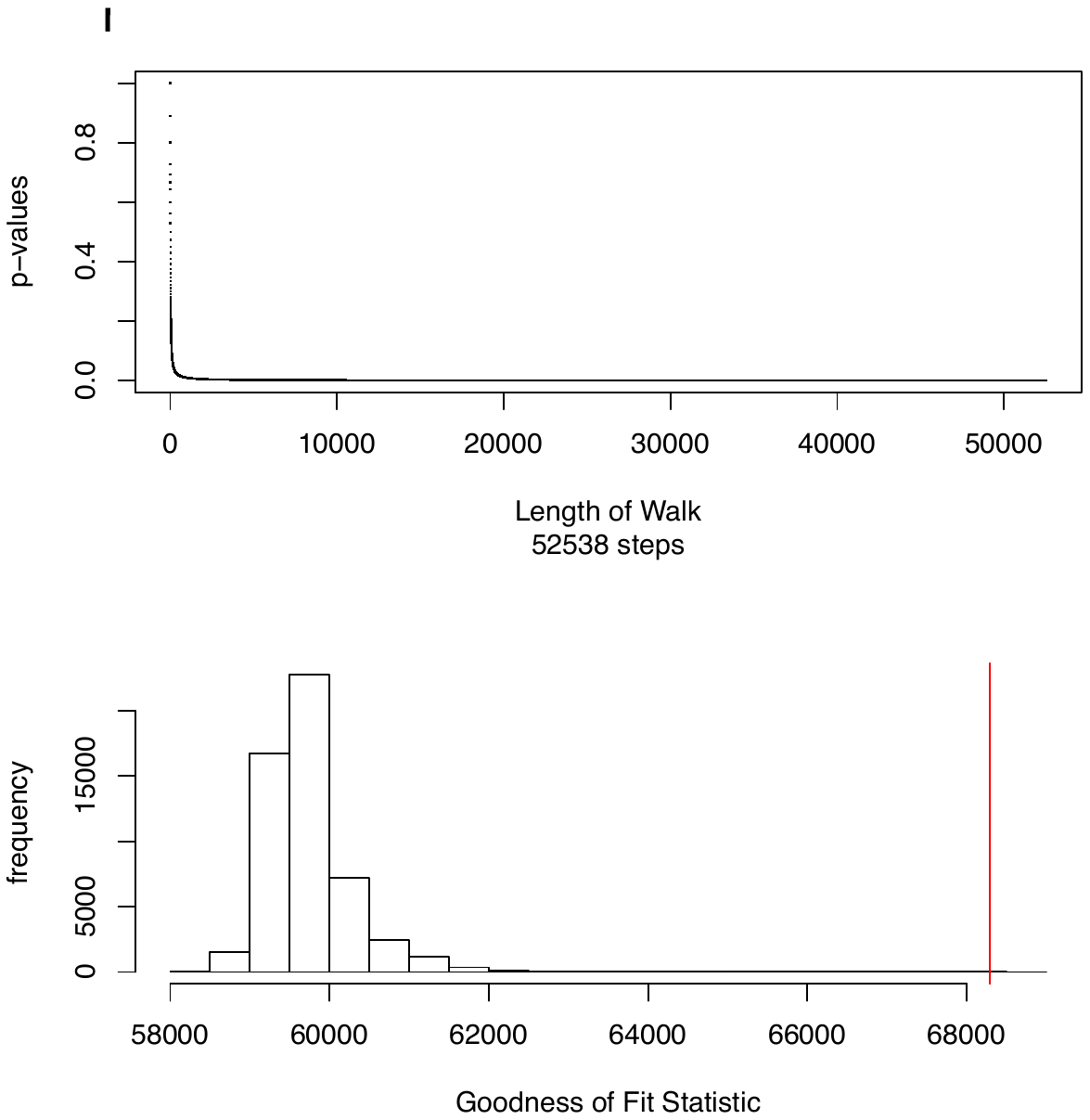}
\caption{Test of model fit of the $\beta$-SBM model to the neuronal mixed network. Block assignment function is given by RegionBlocks. Left: $p$-value quartiles for $3$ iterations on 200,000 steps each. Right: typical $p$-value estimate from one of the iterations.}
\label{fig:betaSBMneuronalRegionBlocks}
\end{subfigure}

\begin{subfigure}[b]{1\linewidth}
\centering 
\includegraphics[width=.35\linewidth]{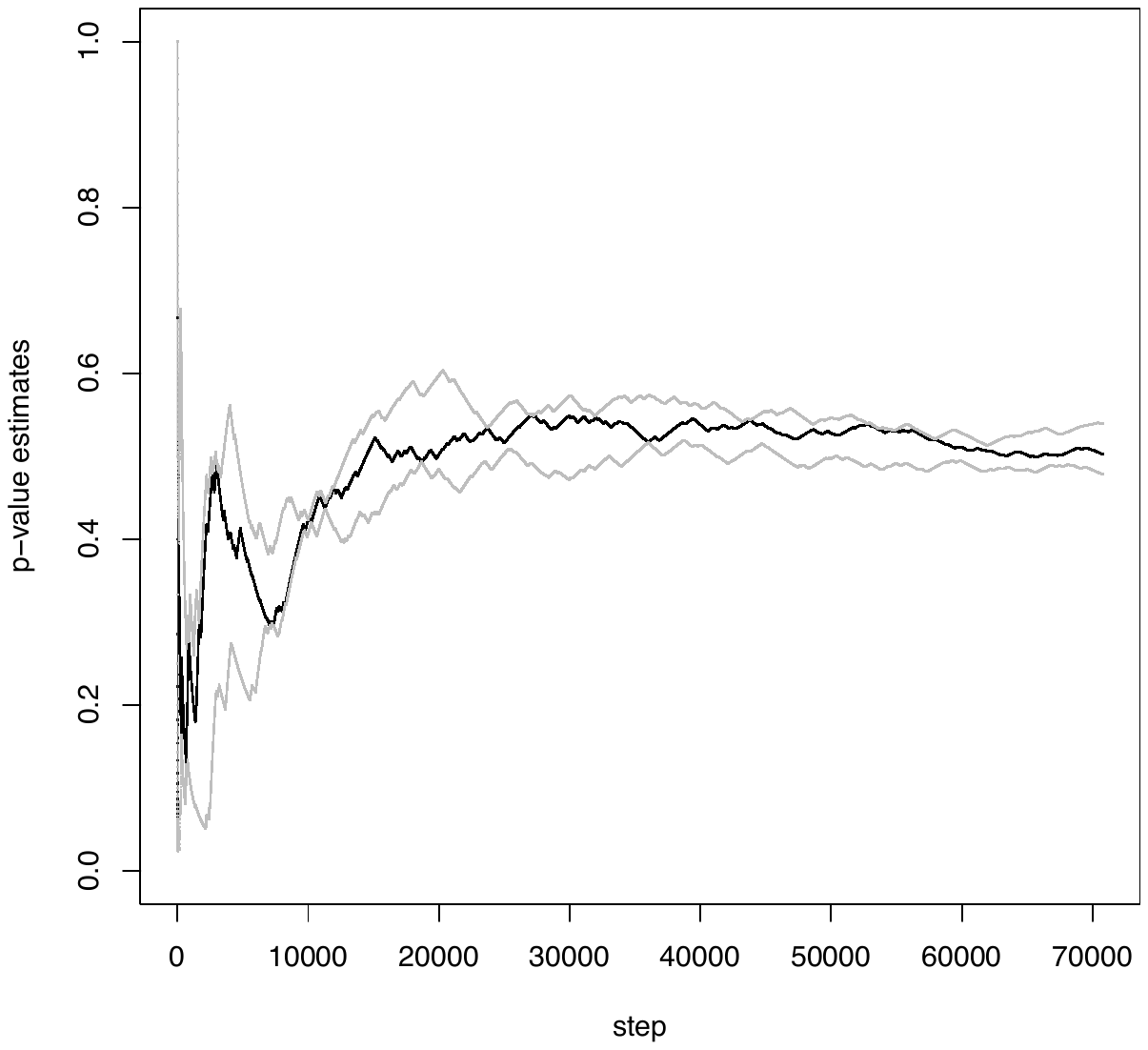}
\centering 
\includegraphics[width=.35\linewidth]{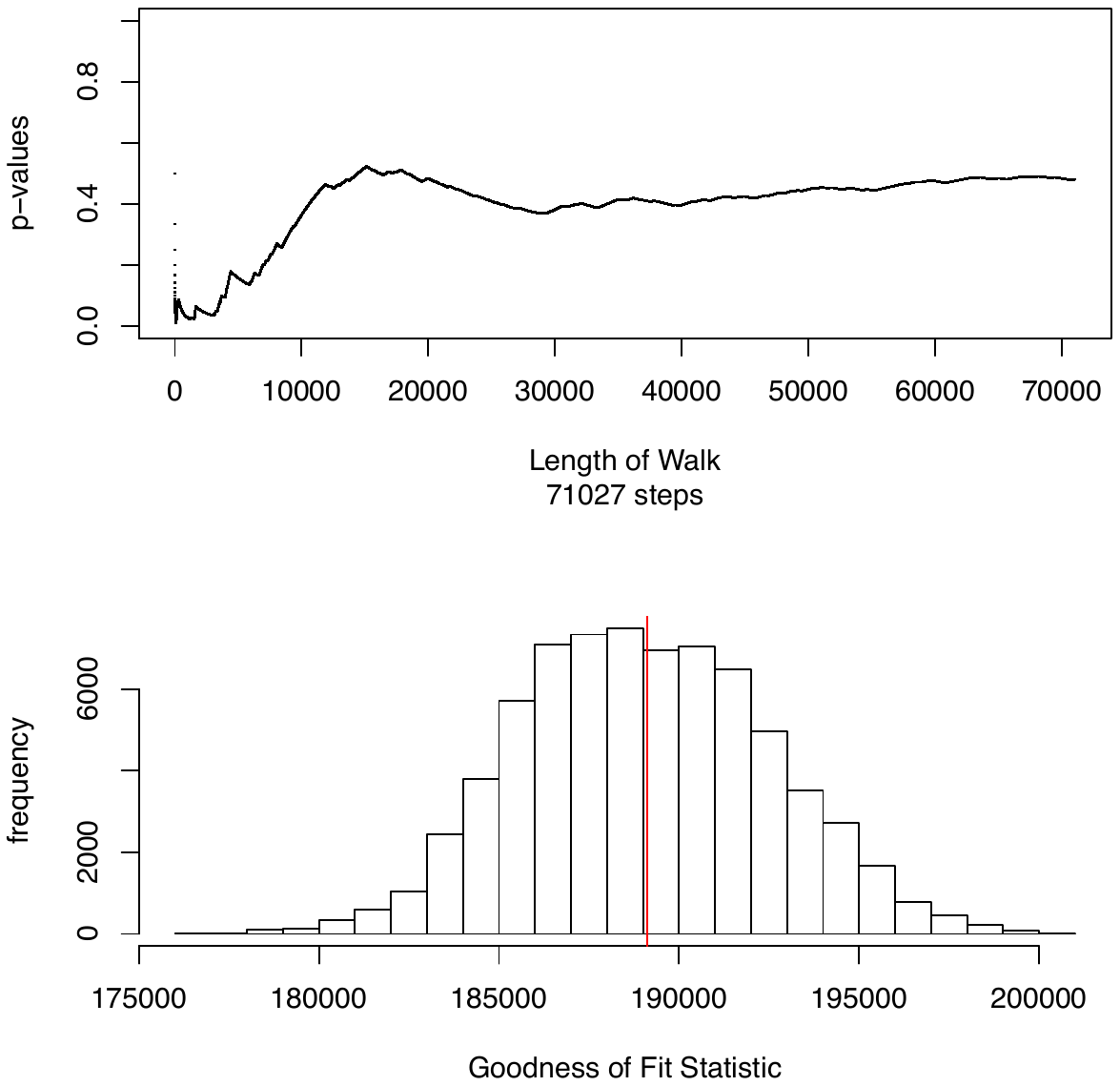}
\caption{Test of model fit for the $p_1$ model with dyad-specific reciprocation effect to the neuronal mixed network consisting of the chemical (undirected) and gap junction (directed) graphs. Left: $p$-value quartiles for $5$ iterations. 
 Right: a typical $p$-value estimate from one of the iterations and the sampling distribution of the goodness-of-fit (chi-square) statistic along with its observed value.}
\label{fig:p1dyadNeuronalDirectedAndUndirected}
\end{subfigure}
\caption{Two models fits on the neuronal mixed network. 
} 
\end{figure}

Next we consider the mixed neuronal network, consisting of the union of the directed graph---the chemical part---and the undirected graph--- the gap junction part.
Since there are three natural neuron groupings -- function, region, and ganglion group -- we add a parameter for possible homophily to the $\beta$-model and again test model fit. For the neuronal mixed network we tested the $\beta$-SBM model with block assignments given by each of the three groupings; since the results were the same for all three, we show it for the region group block assignment. 
Test of model fit of the $\beta$-SBM to the mixed network where the network block assignment is given by {\bf blocks = regionBlocks}: 
 is illustrated in Figure~\ref{fig:betaSBMneuronalRegionBlocks}. Model  rejected at any reasonable significance level, with $p$-value very close to $0$ ($10^{-4}$). 
 Since the $p_1$ model naturally interprets undirected edges as reciprocated, testing dyad-specific reciprocation will help delineate the two types of edges.  Figure~\ref{fig:p1dyadNeuronalDirectedAndUndirected} shows that this model actually fits the data well; whereas the variant with \emph{zero} reciprocation does not (the simulation results, similar to the above, are omitted for length considerations.) 

As we  see in simulation results above, we reject the model for the $p_1$-model with zero reciprocation, which again gives evidence against degree-based edge formation. However, for the mixed network under the $p_1$-model with dyad specific reciprocation, the model is not rejected. 
This could be picking up the signal that some neurons are more likely to be a part of electrical connections as opposed to chemical connections, since the electrical connections are represented as reciprocated edges in the mixed network. 

\subsection{Protein-protein interaction network data}\label{sec:protein} 

The protein-protein interaction network data set is from \cite{AIMC2011} and available from the Plant Interactome Database (\url{http://interactome.dfci.harvard.edu/A_thaliana/}).  The data set is the union of two protein-protein interaction graphs which share a subset of proteins as vertices.  The first protein-protein interaction graph is a literature-curated network consisting of 3,998 directed interactions on 2,160 proteins. The second protein-protein interaction graph is a partial map of the \emph{Arabidopsis thaliana} interactome experimentally constructed in \cite{AIMC2011}.  This second network consists of 2,661 vertices and 5,529 directed edges.  The two networks, $G_1$ and $G_2$ respectively, have 477 proteins in common.  We will consider the union of these two directed graphs, which results in a network with $4, 344$ vertices and $9,449$ edges.  For testing, we will treat edges with one vertex in $V(G_1) \setminus V(G_2)$ and the other in $V(G_2) \setminus V(G_1)$ (regardless of direction) as structural zeros.

\paragraph{Simulation results.} 
As with neuronal network, we also  test the \emph{A. thaliana} protein-protein interaction network for evidence of degree-based edge formation. However, if we test the full network with 4,344 vertices and 9,449 edges, there is a large number of edges that we need to treat as structural zeros since these edges were not tested in the experiment. 
For a data set of this size, that number of structural zeros slows down the algorithm. 
%

\begin{figure}[t]
\begin{subfigure}{.6\linewidth}
\centering 
\includegraphics[width=.8\linewidth]{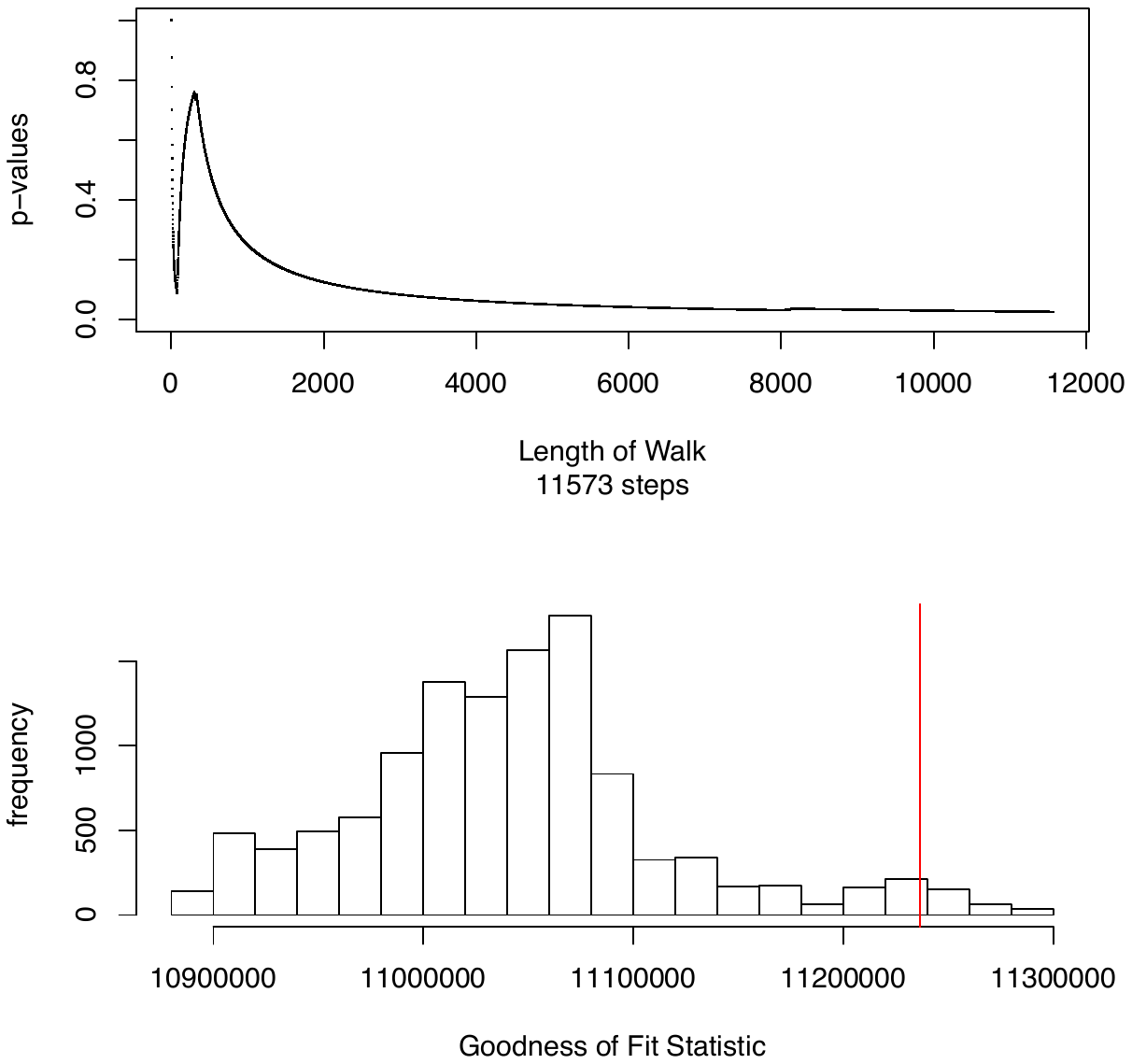}
\caption{A 20,000-step Markov chain 
 for the dyad-specific $p_1$ model test on the protein-protein interaction network with structural zeros. 
 }
\label{fig:proteinUnionRun1short}
\end{subfigure} 
\quad 
\begin{subfigure}{.35\linewidth}
\centering 
\includegraphics[width=1.0\linewidth]{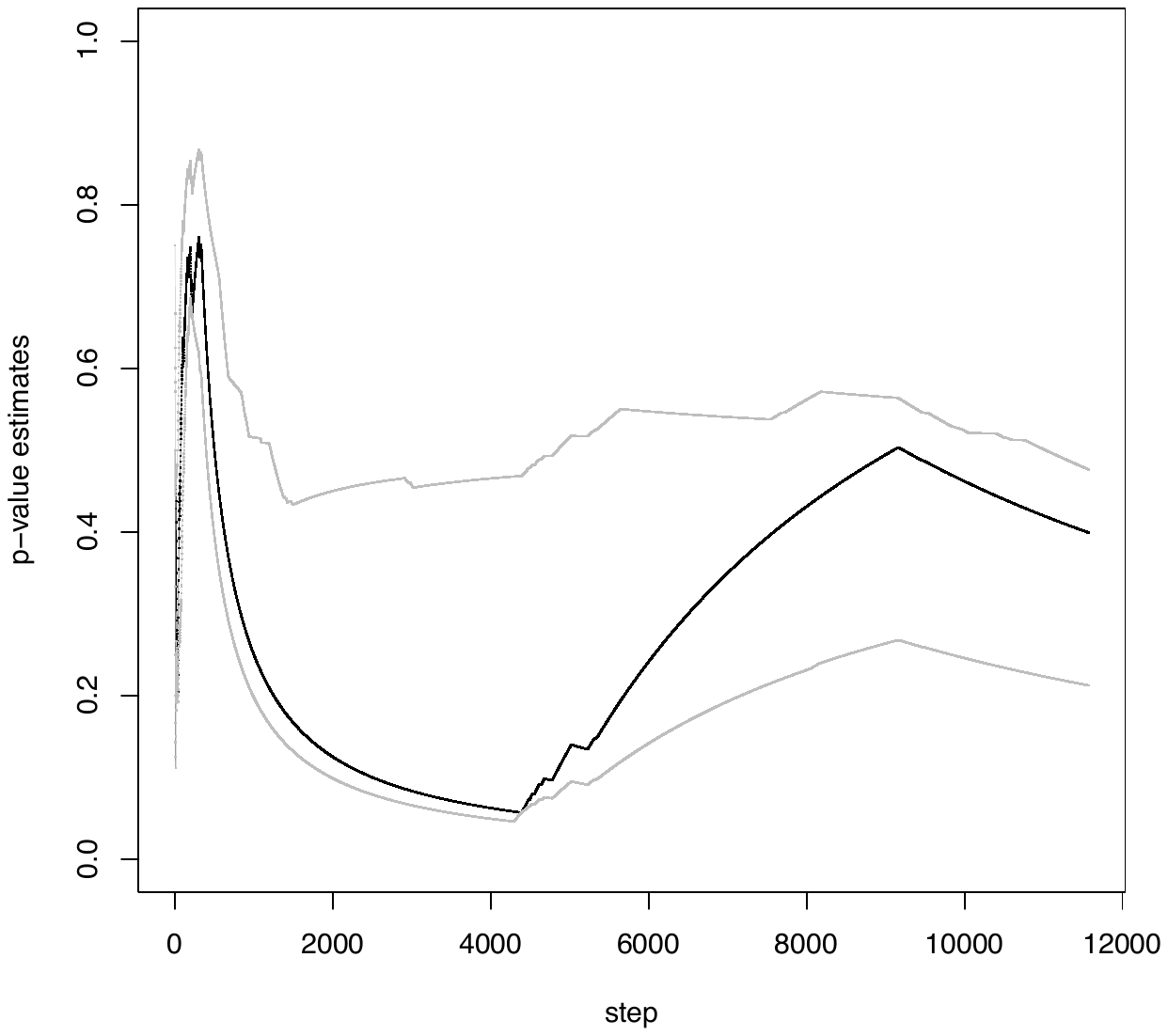}
\caption{$p$-value convergence for the dyad-specific $p_1$ model test on the protein-protein interaction network with structural zeros.}
\label{fig:proteinUnionQuartiles}
\end{subfigure}
\caption{Protein-protein interaction network with a large number of structural zero edges.}
\end{figure}

Thus, we ran the goodness of fit test after  some optimizing, namely, a parallel implementation, small moves tuning parameter  that allows to tune the size of move and make the chain `move' faster, etc.,   and tested the $p_1$ model with dyad-specific reciprocation effect, taking into account the structural zeros. 
Several short simulations completed gave conflicting results, with $p$-values ranging from $0.02$ to $0.2$. 
Due to the prohibitive computation time,  we only ran chains of effective size ~20,000 (steps counted disregarding rejected move proposals), arguably insufficient, and did notice that the chains had not mixed yet. 
A summary of one such run can be seen in Figure~\ref{fig:proteinUnionRun1short}. 
Figure~\ref{fig:proteinUnionQuartiles} illustrates potential red flags in terms of $p$-value converging: three iterations of the model fit test were run in parallel, and produced  different results. A closer look at the simulation reveals that not all the chains have mixed, but the move rejection rate was  low. 

We also  tested the same model without structural zeros on one of the component graphs 
and rejected the model after about $15,000$ steps with $p$-values ranging from $0.0005$ to $0.05$. 

The reader should note that the prohibitive computation time is not imposed   by  how our current implementation handles structural zeros or generation of new graphs. Rather, it is a result of  other computational issues, such as computing the values of the chi-square statistic at each discovered network. For each new graph generated in the fiber, the chi-square computation requires on the order of a few billion small computations, even though we are using the most optimized version of it. We do not yet have a principled approach  to address the challenge of this particular part of the computation to make it more scalable. The reader is referred to Section~\ref{sec:conclusion} for more details.

\subsection{Comparison with existing methods}\label{sec:comparison}

Finally, in this section we compare our results with the goodness-of-fit test implemented in the {\tt R} package {\tt ergm} package \cite{ergm}. Another related goodness-of-fit method includes
 \cite{krivitsky2012exponential}, which describes a Monte Carlo goodness-of-fit test that simulates a statistic of interest from the fitted model and compares it with its observed value, and was more recently formalized as a score test in \cite{SchweinbergerMichael2023DoT}. 

\begin{figure} 
\centering 
\includegraphics[width=1\linewidth]{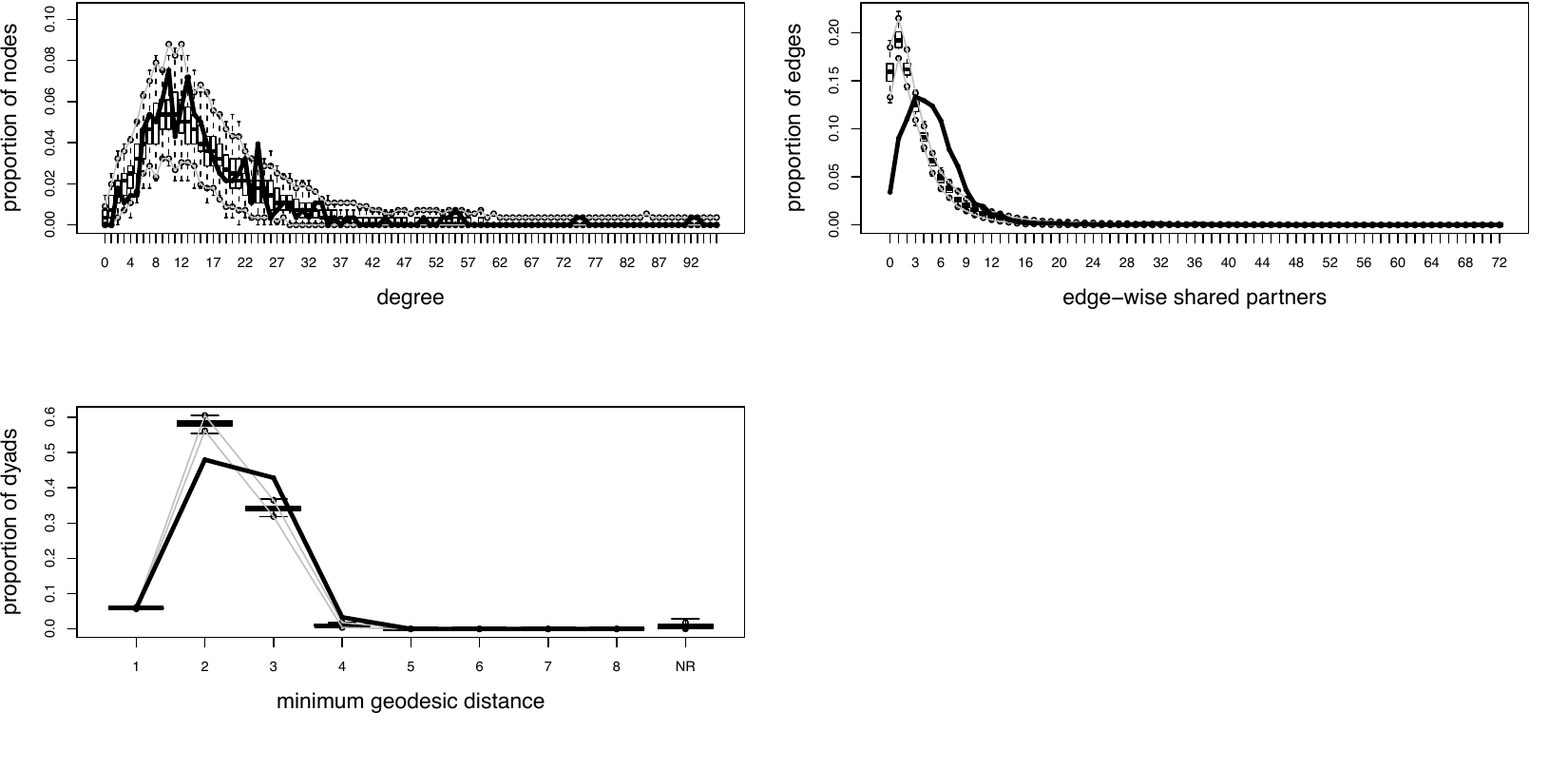}
\caption{Comparison of degree, edge-wise shared partners, and minimum geodesic distance statistics between simulated networks and the mixed, undirected neuronal network.}
\label{fig:gofheuristics}
\end{figure}

For comparison, we will  restrict our attention to the $\beta$--SBM model, which can also be fitted with {\tt ergm}. 
The goodness-of-fit method {\tt gof} implemented in {\tt ergm} is based on \cite{Hunter}.  The method {\tt gof} simulates networks from the fitted model and then computes network statistics of the observed and simulated networks.  In addition to outputting simulated $p$-values, the method also outputs graphical information that  the user can use to visually inspect how the distribution of each statistic for the observed network compares with the distributions of the statistic for the simulated networks.   Here, using the {\tt ergm} package, we fit the $\beta$-SBM model to the mixed, undirected neuronal network with blocks assigned according to ``region," and then run {\tt gof} using the three suggested statistics from \cite{Hunter}: degree, edge-wise shared partners, and minimum geodesic distance. The three plots of these statistics are displayed in Figure \ref{fig:gofheuristics}; the black line represents the statistics of the neuronal network, while the grey lines represent the range for 95 percent of the simulated statistics.  While the degree statistics of the neuronal network are within the range of the degree statistics of the simulated networks, the edge-wise shared partners and minimum geodesic distances of the neuronal network differ greatly from the simulated networks. Thus using these plots, we conclude that the $\beta$-SBM model is a poor fit for the data, a conclusion supported by \S\ref{sec:neuronal}.

We also further experiment with other MCMC algorithms to speed up computations, namely Besag and Clifford's ``parallel method" from    \cite{BesagClifford89generalizedMCsignificance}. In this method, one runs the chain backwards from the observed network $G_0$ for a prescribed number of steps to obtain a network $G_1$. From $G_1$, one then runs $n-1$ independent chains forward to get $n$ simulated random networks that form an approximately independent sample drawn from the uniform distribution on the fiber. While this method worked well for smaller networks, in unoptimized form, it did not scale well for the examples in this paper.

\section{Discussion and next steps}
\label{sec:conclusion}


%

The guiding question of this work is how to perform statistically satisfying goodness-of-fit tests for a family of network models  with good statistical properties and enough flexibility for broad use in applications. 
While non-asymptotic  tests for general ERGMs remain a generally hard problem, we derive  finite-sample tests for model fit for a meaningful subclass of ERGMs called  log-linear ERGMs.  This is done by importing tools from contingency table analysis and adapting them to the combinatorial setting of random graphs, namely,  scalable estimation IPS algorithms and dynamic Markov bases  for sampling from conditional distributions. 
The log-linear setting is a familiar setting for networks and, in particular, this paper is rooted in the connection established in the 1980s by \cite{FW81} and \cite{FienbergWasserman1981categorical}. 

As an application, we test two popular types of biological datasets in network science, a neuronal network and a protein-protein interaction network.  In particular, we test whether the datasets fit several degree-based models, including models with homophily and reciprocation effects. These datasets were chosen since these networks routinely appear as examples of scale-free networks, suggesting degree-based edge formation mechanisms, however, while some authors have suggested exploring biological networks with ERGMs \cite{Saul2007}, \cite{Simpson2011}, up to this point, goodness of fit testing has only been ad hoc.  Thus, this work adds not only to the conversation on goodness-of-fit testing for ERGMs, but also to the conversation in network science focused on understanding the general structure of protein-protein interaction networks and neuronal networks.  In summary, we rejected most of the models that we fitted to these datasets, with the exception of the $p_1$-model with dyad specific reciprocation for the mixed neuronal network containing both the electrical and chemical subnetworks.

\medskip

The largest network we tested was a protein-protein interaction network with 4,344 nodes and 9,449 edges (see Section \ref{sec:protein}). Testing how well this network fit different variants of the $p_1$-model with a large number of structural zeros required a prohibitive, or at least impractical, computation time, which illustrates the limits of the unoptimized way in which we implemented our algorithms in {\tt R}.   One bottleneck in the computation is {\bf not} the graph sampling, rather, it is that the $\chi^2$ statistic is computed at each step in the walk in order to obtain an estimate for the conditional $p$-value.  For example, this computation alone requires finding the entry-wise difference between two matrices of size $4344 \times 4344=18,870,336$.   As fields are moving toward larger and larger datasets, clever computing strategies--or calls to lower-level implementation of such operations--will be needed to handle networks with a million or more nodes.  However, we expect there is ample room for improvement by exploiting current research in combinatorial algorithms and the computer science side of data science.

While more work needs to be done to build an efficient algorithm for testing massive networks, the results of this work illustrate the feasibility of straightforward algorithms that utilize Markov bases for medium-sized datasets. In fact, scalability of algorithms that utilize Markov bases have been an open question in the last decade, since such algorithms traditionally rely on pre-computing the entire Markov basis.  Such computations are not only time and resource intensive but also produce mostly inapplicable moves since, by definition, these bases are independent of the observed data \cite[Problem 5.5]{DobraEtAl-IMA}; cf. \cite{F-review}, and are not specialized to any particular fiber of the model. 
 With this issue in mind, various proposals have been made to make exact conditional tests for tables scalable to larger data or more complex models. For example, 
\cite{Dobra2012} proposes a general dynamic approach to constructing applicable local moves for marginal table models and proves that they can be utilized to cover the entire fiber. Meanwhile, \cite{Hara2012} use a Poisson-size combination of the smallest possible set of moves to explore fibers of discrete logistic regression models, and \cite{UhlerEtAl-IsingModel} use another small subset of a Markov basis for testing the Ising model on a large biological dataset.  Each of these methods extends the use of Markov bases to larger and larger datasets: the approach in \cite{Dobra2012} was demonstrated on tables up to 256 cells, \cite{Hara2012} show their method approaches its limits on $10 \times 10 \times 10$ tables, and \cite{UhlerEtAl-IsingModel} are able to use their method on a biological dataset of size $800 \times 800$.  In this paper, we are able to obtain good results for tables up of size $2661 \times 2661 \times 2 \times 2$ (networks with 2661 vertices) and show that 
our method slows down due to a goodness-of-fit statistic computation
 on tables of size $4334 \times 4334 \times 2 \times 2$ (networks with 4334 vertices). 

\smallskip 
One immediate benefit of using a contingency table representation and Markov bases is the ease of   generalization of the goodness of fit test to models for weighted networks, represented by graphs with positive integer weights on the edges. This can be achieved by simply removing the $0/1$ table entry sampling restriction; the resulting estimation algorithms are not affected, and the sampling algorithms based on Markov bases, in fact, become simpler because they do not require the additional step of checking for the graph being simple. Due to this simplification, we expect the sampling algorithm's mixing time and scalability to improve in this setting. 
In regards to the mixing time of the proposed algorithms in this manuscript, we note that \cite{DillonMS16}  showed that the dynamic Markov bases algorithm from \cite{GPS16} is rapidly mixing on many fibers, since it contains the simple switch chain well-known in the graph theory literature \cite{KannanEtAl99}.  Although  herein we propose many more chains than \cite{GPS16}, most of these chains also contain the simple switch chain, which provides ample evidence that the algorithms will mix rapidly on most fibers, even if a formal mixing time analysis is outside the scope of this paper.  

On a final note, if it is known that the data does not support dyadic independence, analysis using more complicated ERGMs outside of the log-linear ERGM class would be desirable, though tests of model fit is a looming question. 
 In such cases, the sufficient statistic is no longer linear in terms of the dyads, and the geometry and combinatorics of the fibers of the resulting model (that is, graphs with the fixed value of the sufficient statistic) is poorly understood. 
 While there exist models that do not assume dyadic independence, ranging from the very simple edge-triangle model \cite{Strauss, ParkNewmanET04, ParkNewmanET05} 
 to those using global summary statistics \cite{shellERGM}, it is not clear how to use any available tools for testing the dyadic independence assumption on a given network from a statistical point of view. One may turn to estimating graphical models on dyads such as \cite{FS86} or more recent general models from \cite{AleKayvanGraphicalNtwks}, 
however for these models, testing goodness of fit is an open problem as well. 


\bibliography{DynMarkP1,BioNetworks,AlgStatNtwks,GoFandSBM,discreteMethodsAlgStatSurveyPARTIAL,generalNetworks}

\appendix                                     
\section{Parameter hypergraphs}\label{sec:appendix:hypergraphs}

In this appendix, we define and give several examples of a combinatorial structure that encodes any log-linear model: the \emph{parameter hypergraph}. 
One should think of this discrete structure object as a schematic representation of the model parametrization, as   it encodes the way in which parameters of the model interact. For example in the $\beta$-model, log-odds of each edge probability $p_{ij}$,  $\beta_i+\beta_j$, is encoded by grouping the two parameters $\{\beta_i,\beta_j\}$ together. Each such set represents one edge in the parameter hypergraph.   
In mathematically precise terms, 
a \emph{hypergraph} $\mathcal H = (\mathcal V, \mathcal E)$ is an ordered pair where $\mathcal V$ is a set of elements called \emph{vertices} and $\mathcal E$ is a set of non-empty subsets of $\mathcal V$ called \emph{hyperedges}, or simply \emph{edges}. Note that a graph is an instance of a hypergraph where all edges are subsets of $\mathcal V$ of size 2.

The general setup is that  the vertices  of the parameter hypergraph are the model parameters, while the edges are  sets  of parameters that appear together in any one of edge probability (or, equivalently, their log-odds). 
In particular, the edge sets for the degree-based models below should be 
compared to the equations in Section 
\ref{sec:knownmodels}.


\subsection{Combinatorics of $\beta$-model}
The parameter hypergraph $\mathcal H = (\mathcal V, \mathcal E)$ of the $\beta$-model on $\mathcal G_n$  has vertex set 
$$\mathcal V = \{ \beta_i \ : \ 1 \leq i \leq n\}$$
and edge set 
$$ \mathcal E = \{ \{\beta_i, \beta_j\} \ : \ 1 \leq i < j \leq n\}.$$ 
Notice that for the $\beta$-model, the parameter hypergraph is the complete simple graph on $n$ vertices.

\subsection{Combinatorics of the $p_1$ model with zero reciprocation}\label{sec:hypP1zero}

The parameter hypergraph of the $p_1$-model with zero reciprocation on $\mathcal G_n$ is $\mathcal H = (\mathcal V, \mathcal E)$ where
$$\mathcal V = \{ \alpha_i, \ \beta_j \ : \ 1 \leq i \leq n, \  1 \leq j \leq n\} \text{ and } \mathcal E = \{ \{\alpha_i, \beta_j\} \ : \ 1 \leq i, j \leq n, \ i\neq j\}.$$ 
Notice that by restricting to the edges of size two in $\mathcal E$, we obtain the complete bipartite graph $K_{n,n}$ with the edges $\{\alpha_i, \beta_i\}$ removed.

A careful reader may notice the absence of the original parameters $\lambda_{ij}$. These are normalizers added to the model parametrization in order to ensure that each dyad is observed in exactly one of the four states. 
Combinatorially, In this particular model, these parameters are redundant as they do not change the underlying structure from which Markov moves for sampling are derived, so we choose to sample from the simplified parameter hypergraph instead. 

\subsection{Combinatorics of the $p_1$ model with constant reciprocation}\label{sec:hypP1const}

The parameter hypergraph of the $p_1$-model with constant reciprocation on $\mathcal G_n$ is $\mathcal H = (\mathcal V, \mathcal E)$ where
$$\mathcal V = \{ \alpha_i, \ \beta_j, \rho \ : \ 1 \leq i \leq n, \  1 \leq j \leq n\} \text{ and } \mathcal E = \{ \{\alpha_i, \beta_j\}, \{\alpha_i, \alpha_j, \beta_i, \beta_j, \rho\} \ : \ 1 \leq i, j \leq n, \ i\neq j\}.$$ 
Notice that, as in the zero reciprocation case, by restricting to the edges of size two in $\mathcal E$, we obtain the complete bipartite graph $K_{n,n}$ with the edges $\{\alpha_i, \beta_i\}$ removed.

\subsection{Combinatorics of the $p_1$ model with dyad-specific reciprocation}

The parameter hypergraph of the $p_1$-model with constant reciprocation on $\mathcal G_n$ is $\mathcal H = (\mathcal V, \mathcal E)$ where
$$\mathcal V = \{ \alpha_i, \ \beta_j, \rho \ : \ 1 \leq i \leq n, \  1 \leq j \leq n\} \text{ and } \mathcal E = \{ \{\alpha_i, \beta_j\}, \{\alpha_i, \alpha_j, \beta_i, \beta_j, \rho_i, \rho_j\} \ : \ 1 \leq i, j \leq n, \ i\neq j\}.$$ 
As in the previous two $p_1$ models, restricting to the edges of size two in $\mathcal E$, we obtain the complete bipartite graph $K_{n,n}$ with the edges $\{\alpha_i, \beta_i\}$ removed.  Additionally, notice the induced hypergraph on $\mathcal V_{\rho} = \{ \rho_i \ : \ 1 \leq i \leq n\}$ gives us the complete graph on $n$ vertices.

\section{The conversion step}\label{sec:appendix:table_to_hypergraph}

Step~\ref{step:convert} of Algorithm~\ref{alg:NextMove} is a subroutine to convert a dyad classification table $v$ to a multihypergraph $\mathbb R$. 
The log-linear model matrix $A$ is the incidence model of the parameter hypergraph. This means that the rows of $A$ are indexed by vertices (model parameters). 
There are as many columns of $A$ as there are cells in the dyadic classification table $u$ of a network $g$. 

The conversion then is as follows. Each nonzero entry in a cell $e$ of a given subtable $v$ is encoded as an edge of the parameter graph corresponding to the column $A_e$. 

For example, in the $p_1$ model with constant reciprocation (Section~\ref{sec:hypP1const}), an entry $m\in\mathbb Z_{\geq 0}$ in the cell representing the dyad $\{i,j\}$ in state $i\rightarrow j$ is converted into  $m$ copies of the hyperedge $\{\alpha_i,\beta_j\}$. Entry $m$ in the cell representing the dyad $\{i,j\}$ in the state $i\leftrightarrow j$ is converted into  $m$ copies of the hyperedge $\{\alpha_i,\beta_j,\alpha_j,\beta_i,\rho\}$.

\end{document}